\documentclass[twocolumn,trackchanges]{aastex701}
\received{October 30, 2025}
\submitjournal{ApJ}

\begin{document}

\title{The Universality of Dark Matter Density Profiles for Milky Way Analog Galaxies}

\author[orcid=0009-0001-6634-1065,gname=Maria Clara, sname=Cavalcante Siviero]{Maria Clara Cavalcante-Siviero}
\affiliation{Brazilian Center for Research in Physics}
\email[show]{mariaclara@cbpf.br}

\author[orcid=0000-0003-3153-5123,gname=Karín, sname=Menéndez Delmestre]{K. Menéndez-Delmestre} 
\altaffiliation{Valongo Observatory}
\affiliation{Federal University of Rio de Janeiro, Ladeira de Pedro Antônio 43, Rio de Janeiro, RJ 20080090, Brazil}
\email{kmd@astro.ufrj.br}

\author[orcid=0000-0001-5071-1343,gname=Pedro,sname=Beaklini]{P. P. B. Beaklini}
\affiliation{National Radio Astronomy Observatory, Array Operations Center, Socorro, NM 87801, USA}
\email{pbeaklin@nrao.edu}

\author[orcid=0000-0003-2374-366X,gname=Thiago, sname=Signorini Gonçalves]{T. S. Gonçalves} 
\altaffiliation{Valongo Observatory}
\affiliation{Federal University of Rio de Janeiro, Ladeira de Pedro Antônio 43, Rio de Janeiro, RJ 20080090, Brazil}
\email{}

\author[orcid=0000-0003-1683-5443,gname=Davi,sname=Rodrigues]{D. C. Rodrigues}
\affiliation{Physics departament, Federal University of Espírito Santo, Vitória, ES, Brazil}
\email{}

\author[orcid=0009-0005-1424-6604,gname=Nathan,sname=de Isídio]{N. G. de Isídio}
\affiliation{European Southern Observatory, Garching, Germany}
\email{}

\author[orcid=0000-0001-5539-0008,gname=Amanda,sname=Araujo-Carvalho]{A. E. Araujo-Carvalho}
\altaffiliation{Valongo Observatory}
\affiliation{Federal University of Rio de Janeiro, Ladeira de Pedro Antônio 43, Rio de Janeiro, RJ 20080090, Brazil}
\email{}

\begin{abstract}

The structure, extent, and mass of the Milky Way's (MW) dark matter (DM) halo are observationally challenging to determine due to our position within the Galaxy. To overcome this limitation, we study a combined sample of 127 MW analogs from the IllustrisTNG-50 cosmological simulation with observations of 11 nearby galaxies. Using both spatial and spectral high-resolution data from VLA and GMRT telescopes, we employ the 3D-Barolo algorithm to derive precise kinematic maps and rotation curves (RCs). To perform a careful analysis of the stellar component, we use Spitzer mid-IR imaging at 3.6 and 4.5 $\mu m$. We decompose the RCs into their different mass components, enabling the construction of a DM radial profile for each galaxy. By using a MCMC-based routine, we account for the DM contribution for the observed RCs. For our simulated sample, we obtain DM radial profiles directly from the TNG50 database. We probe for the universality of the DM profiles by deriving and comparing the equivalent local DM density (LDMD), a critical parameter linked to DM direct detection experiments on Earth. We calculate the DM density at the corresponding location of the Sun in each of the analogs. Our analysis yields a final LDMD range of $0.17-0.46~GeV~cm^{-3}$. Finally, by leveraging our mass estimates ($M_{200}$, $M_{gas}$ and $M_{\star}$), we contextualize our findings with the efficiency of star formation in MW analogs and with the diversity of galaxies inhabiting similar halo masses.

\end{abstract}

\section{Introduction}

According to the $\Lambda$CDM cosmological model, baryonic matter constitutes only $\sim 15\%$ of the total matter content in the universe \citep{planck}, with the remaining 85$\%$ attributed to the unseen component known as dark matter (DM). Despite many astrophysical evidences for the existence of DM, from galaxy rotation curves \citep[RCs;][]{rubinford1980,bosma1981,sofuerubin2001,mancera2022,DiTeodoro2022} to the large scale structure of the Universe \citep{Davis1985,nfw1996}, the nature of DM still eludes us.
The leading theoretical candidates are Weakly Interactive Massive Particles \citep[WIMPs,][]{STEIGMAN1985375}. These particles naturally arise in various extensions of the Standard Model of particle physics and, within the $\Lambda$CDM paradigm, successfully satisfy cosmological constraints from the early universe. 
Specifically, cold DM (non relativistic at all times) is required to correctly form the observed cosmic structure \citep[e.g.][]{riess1998,spergel2007}, as warmer, lighter particles would suppress the formation of small-scale halos where galaxies are formed \citep{Tremaine1979,White1983}.
In agreement with the early suggestion by \citet{Peebles1982} on the nature of cold DM, we assume that it might consist of a new, weakly interacting, neutral particle.
Direct detection experiments on Earth aim to capture any phenomenological effect from DM particles interacting with standard matter. These experiments search for faint signals, such as phonons (heat), scintillations (light), or ionization (charge), generated by a hypothetical collision (recoil) between a DM particle and a target nucleus \citep[see][for a extended review]{Undagoitia_2015}. The expected interaction rate for any given DM model is directly proportional to the DM density at a specific galactic address. Consequently, precise estimates of the local DM density (LDMD) are crucial for interpreting the results and setting constraints on particle physics parameters in experiments like those conducted by the XENON \citep{Aprile_2017} and LUX-ZEPLIN \citep[LZ;][]{LZ,Aalbers_2023,Akerib_2020} collaborations.

Historically, two primary approaches have been employed to estimate the LDMD in the MW. The first one relies on global mass modeling of the Galactic RC (e.g., \citealt{Weber_2010, Catena_2010,iocco2011,Sofue_2013,Sofue2015}). This method constructs mass models for the baryonic components (e.g., \citealt{Salucci_2010,Catena_2010,Nesti_2013,McMillan2017,deisídio2024}) and fits a DM halo profile based on overall kinematics. While these models are often well motivated and can achieve uncertainties below $\sim10\%$ \citep{Catena_2010, IoccoPato_2015}, they are fundamentally limited by their dependence on specific, and often debatable, assumptions regarding the shape of the DM halo (e.g., spherical versus triaxial; \citealt{reid2014,Palau_2023}) and the mass-to-light ratios (M/L) of the baryonic components. The second approach makes use of local kinematic measurements from (mostly) stellar tracers \citep{Kapteyn_1922,Oort1932,Bovy2012,Bienayme2014, zhang2013}. By inferring the gravitational potential from vertical motions of stars above the Galactic midplane, one can directly determine the LDMD without assuming a global halo model. However, this method presents its own challenges: many studies have found results consistent with $\rho_{DM}\approx0$ even at the $3\sigma$ level, unless strong assumptions are made about the dynamics of the tracer population \citep{IoccoPato_2015}.
More recently, a third approach has been put forward to directly measure LDMD through stellar accelerations within the MW \citep[e.g.,][]{Silverwood:2018qra, Chakrabarti_2020}. 
Direct measurement of these accelerations is achievable through high-precision techniques such as pulsar timing and the study of eclipsing binary stars \citep{Chakrabarti_2021, Chakrabarti_2022}. These approaches are particularly promising for probing DM sub-structure and will be complemented by next-generation time-domain facilities like the James Webb Space Telescope and Roman Space Telescope.

Recent work by \citet{deisídio2024} points to the interest of studying external MW analogs to provide independent estimates of the LDMD. If DM is a universal component, its properties should be consistent across dynamically similar galactic environments. To date, only a handful of nearby galaxies have been analyzed for this purpose, with \citet{deisídio2024} presenting a sample of just five systems. They introduce a proof-of-concept methodology, probing that baryonic mass modeling and RC decomposition approach can yield reliable estimates of the LDMD in MW-like galaxies. Despite the limited sample size, \citet{deisídio2024} has successfully established external galaxies as valid laboratories for inferring the DM distribution of the MW itself. However, the lack of substantial statistics underscores the need for larger, more comprehensive samples. 
From a simulation perspective, a recent analysis by \citet{pillepich2024} demonstrates that selecting MW analogs from the IllustrisTNG-50 cosmological simulation can effectively complement observational studies.
These simulations provide access to a diverse population of galactic environments hosted by DM halos of similar mass, enabling systematic investigations of galaxy properties.

In this paper, we explore the DM density profiles for a combined sample of observed and simulated MW-analog galaxies. 
We combine an increasing number of observed galaxies with simulated halos, as a means to expand the breadth of galaxy properties with DM halos similar to our own. 
These external systems may provide reliable estimates for the MW's LDMD, thereby contributing to the increasingly narrow parameter space targeted by terrestrial direct detection experiments. 
This work not only assesses the mass components across a much broader sample of MW-like galaxies, but explores the connection between galaxies and their halos, revealing an underlying universality of the DM density profiles.

The remainder of this paper is organized as follows.
Section~\ref{sec:obs} presents the observational data, including both newly acquired radio interferometric and archival infrared imaging. In Section~\ref{sec:tng50}, we present the sample selection and data from the IllustrisTNG-50 cosmological simulation. The methodology for kinematic modeling is detailed in Section~\ref{sec:methods}. Section~\ref{sec:results} presents our results and discussion. Finally, we summarize our conclusions in Section~\ref{sec:conclusion}. We adopt a $\Lambda$CDM cosmology of plane geometry with $H~=~67.74~[km~s^{-1}~Mpc^{-1}]$, $\Omega_{\Lambda,0}~=~0.6911$, $\Omega_{m,0}~=~0.3089$, $\Omega_{b,0}~=~0.0486$, $\sigma_{8}~=~0.8159$ and $n_{s}~=~0.9667$ \citep{planck}.

\section{OBSERVED MW ANALOGS AND DATA} 
\label{sec:obs}

In this work, we make use of HI kinematic data and stellar mass maps to construct and decompose RCs. To ensure the deepest coverage of stellar mass content in our sample, we select MW analog galaxies from the Spitzer Survey of Stellar Structure in Galaxies \citep[S$^4$G;][]{sheth2010,Munoz_Mateos_2013,querejeta2015,Watkins2022}, the and most homogeneous imaging for the largest sample of nearby galaxies at 3.6 and 4.5 $\mu m$, down to and equivalent mass surface density of $1~M_{\odot}~pc^{2}$. For the HI gas dynamics we performed observations with the Karl G. Jansky Very Large Array (VLA) and the Giant Metrewave Radio Telescope (GMRT). The kinematic analysis is based on a combination of these acquired interferometric radio data.

\subsection{Observations}\label{subsec:vla}

\subsubsection{Sample selection}\label{subsec:sample} 

\begin{figure}
    \centering
    \includegraphics[width=1.1\linewidth]{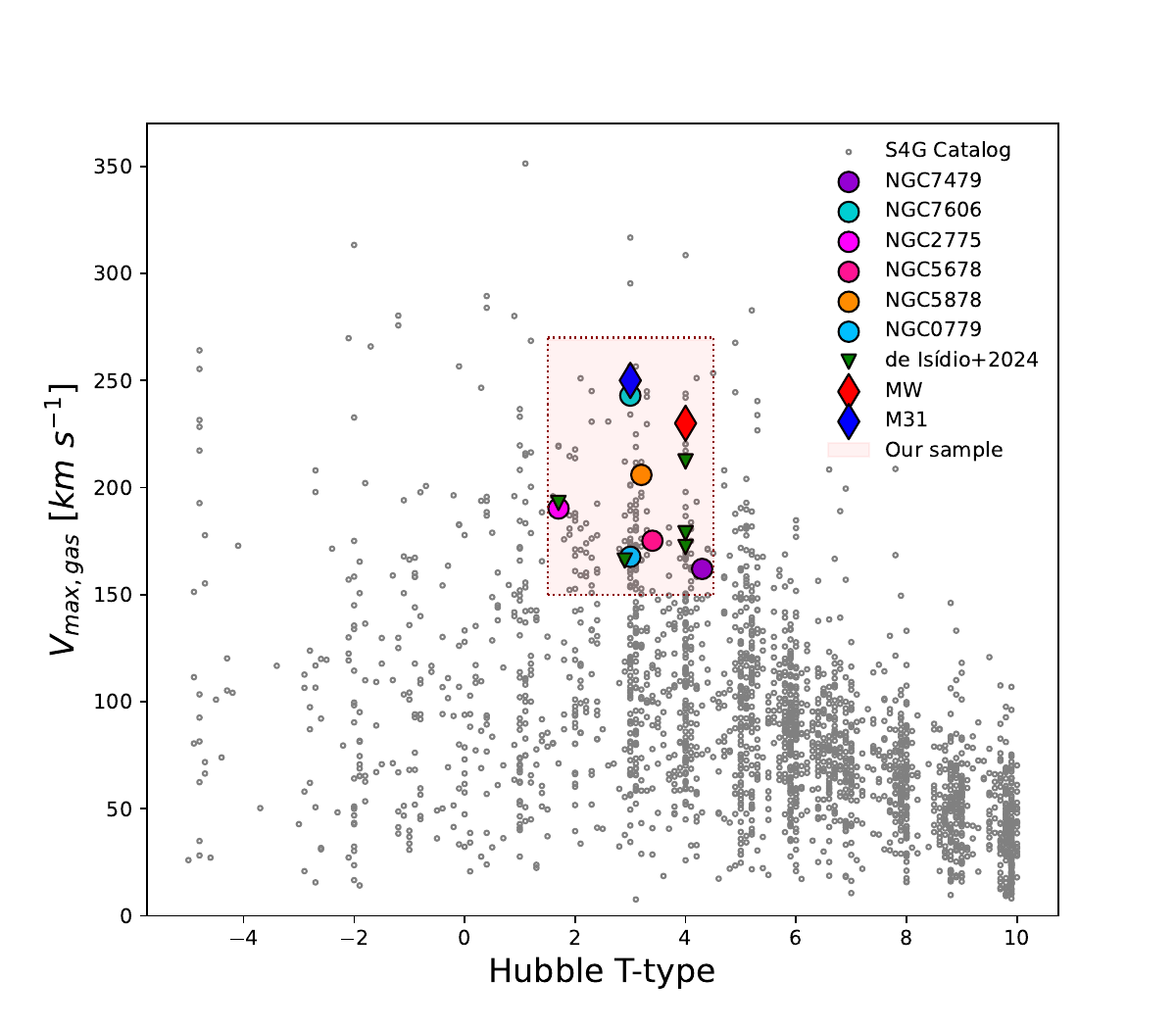}
    \caption{Sample selection from the S$^4$G catalog. Gray points represent the full S$^4$G sample. The light red shaded region indicates our selection criteria in the parameter space. Selected MW analog galaxies are highlighted with colored circles (this work) and green triangles \citep{deisídio2024}. The Milky Way \citep[MW;][]{reid2016} and Andromeda \citep[M31;][]{Carignan_2006} are shown for comparison as red and blue diamonds, respectively.}
    \label{fig:s4gsample}
\end{figure}

We aim to investigate the dynamics and mass distribution in MW-like galaxies. For this, we base our selection on kinematic and morphological criteria: adopting a circular velocity at the Solar circle of $229 \pm 0.2~km~s^{-1}$ \citep{Eilers_2019} and a maximum HI velocity range of $200-250~km^{-1}$ for the MW \citep{reid2016}, we choose galaxies with maximum HI velocities between 150 and 280 $km~s^{-1}$. This ensures that our sample comprises galaxies with halo masses comparable to that of the MW \citep[e.g.,][]{Posti_2019}.
Considering that the MW is classified as SBb/c with a Hubble T-type in the range of 2-3 \citep{Hou_2014}, we select galaxies within the range 1.5-4.5, extending the classification of the MW to include a broader range of late-type spirals (see Figure \ref{fig:s4gsample}). 
We further restrict our sample to include galaxies suitable for reliable RCs, with inclinations between $35^{\circ}$ and $75^{\circ}$.

HI data for five of these systems were already presented in \citet{deisídio2024}, based on VIVA \citep{chung_2009} and THINGS \citep{Walter_2008} surveys. We expand the sample and complement their analysis with new VLA and GMRT HI observations for six new MW analogs. This results in a final observed sample of 11 galaxies. Mid-infrared images from S$^4$G are available for all selected galaxies.

\subsubsection{VLA observations}\label{subsec:vla}

We observed four MW analogs (see Table \ref{tab:sources}) using the VLA array in C configuration at L-band ($1-2~\mathrm{GHz}$) during March 2024, totaling $\sim 17.3$ hours (Proposal ID: VLA/24A-421; PI: K. Menéndez-Delmestre). The largest angular scale in C configuration is $970~\mathrm{arcsec}$ ($16.6~\mathrm{arcmin}$) and sets an upper size limit for detectable extended emission in this interferometric setup. All targets in our sample satisfy this constraint, with $D_{25}$ diameters $\leq 14~\mathrm{arcmin}$. 
We achieved a spectral resolution ranging from $3$ to $13~\mathrm{km~s^{-1}}$, by binning the requested resolution of $6~\mathrm{km~s^{-1}}$ in eight channels. The observations provided a spatial resolution of $15–25\arcsec$, using 27 antennas with a maximum baseline of $3~\mathrm{km}$.

We performed data reduction and calibration using the \textsc{Common Astronomy Software Applications} package \citep[][version 6.6.6]{CASA}. The VLA pipeline in \textsc{CASA} was used to apply standard bandpass, phase, and complex gain calibrations to the data. We also conducted continuum subtraction using line-free channels from the beginning and end of the observed cubes to isolate the HI line emission. 
We performed self-calibration for each VLA scheduling block using field sources to improve the signal-to-noise ratio in the final data cubes.

The achieved RMS noise levels are $5.7\cdot10^{-4}~\mathrm{Jy~beam^{-1}}$ for NGC~7479 and $4.6\cdot 10^{-4} ~\mathrm{Jy~beam^{-1}}$ for NGC~2775. Due to a shorter on-source integration time of only 2.7 hours, NGC~7606 exhibits a higher noise level of $1.4\cdot 10^{-3}~\mathrm{Jy~beam^{-1}}$. All data cubes were imaged using the \textsc{TCLEAN} task in CASA with \textsc{wproject} gridder. We employed \textsc{natural} weighting for NGC~7479 and NGC~2775, and \textsc{Briggs} weighting (robust=0.4) for NGC~7606. The deconvolution was performed with \textsc{hogbom} algorithm for NGC~7479 and \textsc{multiscale} algorithm for NGC~7606 and NGC~2775. 
Figure~\ref{fig:observed} presents the resulting HI integrated intensity maps (moment 0) - generated by summing flux across all channels with detected emission - and the corresponding line spectrum for each source.

\subsubsection{GMRT observations}\label{subsec:gmrt}

The observations with the GMRT telescope were conducted in April-July 2023 totaling $\sim40$ hours (Proposal Code: 44\_006; PI: K. Menéndez-Delmestre). This included 10 hours of on-source integration time per target for each of the four sources. Data were acquired in Band-5 (L-band; 1000–1450 MHz). Self-calibration procedure could not be applied to the GMRT observations due to the absence of suitable calibrator sources within the FOV of our target galaxies. The final data cubes reached a spatial resolution of $3-4\arcsec$ and a spectral resolution of $3-4~\mathrm{km~s}^{-1}$. Individual values for each galaxy are provided in Table~\ref{tab:sources}.

\begin{figure*}
    \centering
    \includegraphics[width=0.3\linewidth]{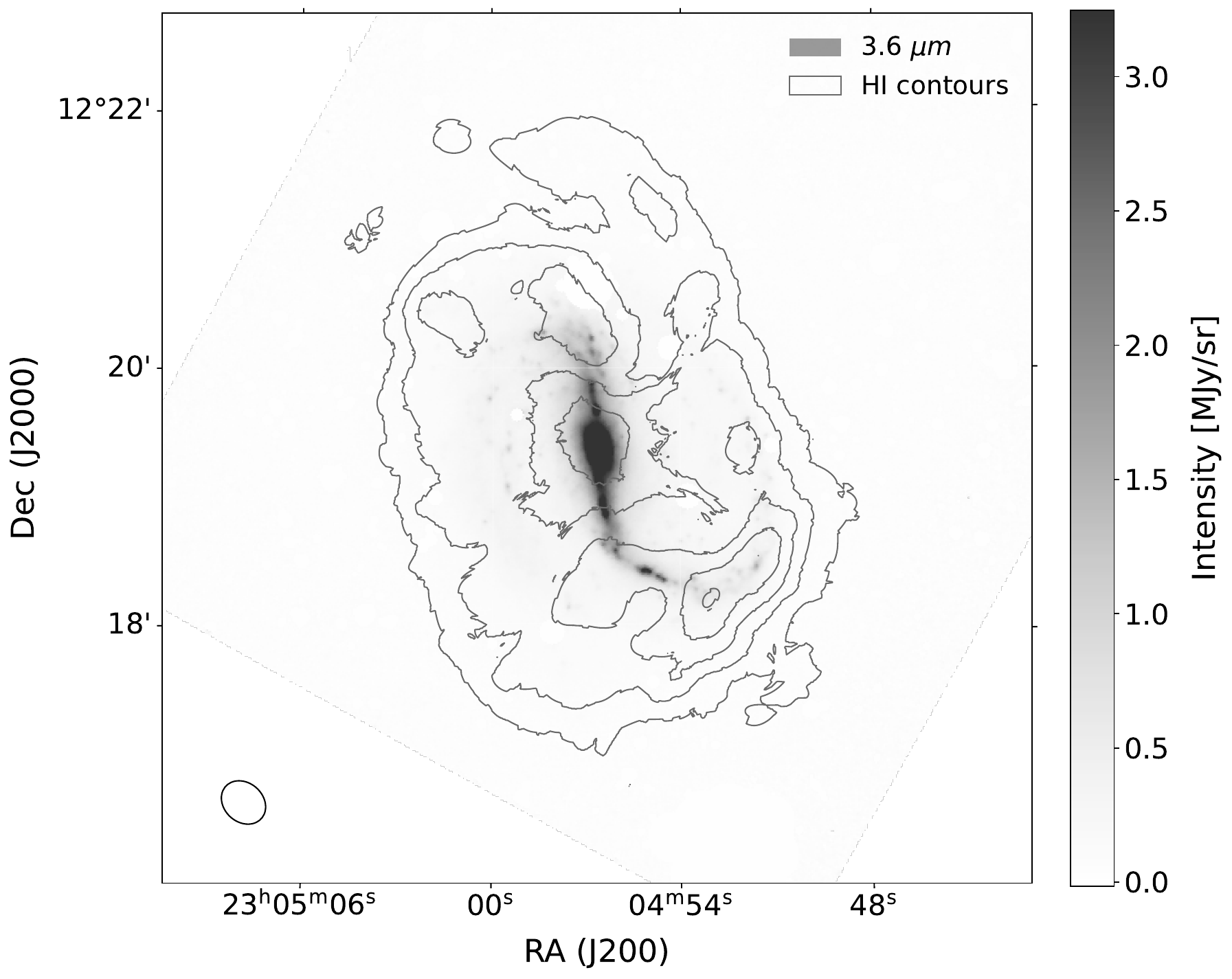}
    \includegraphics[width=0.3\linewidth]{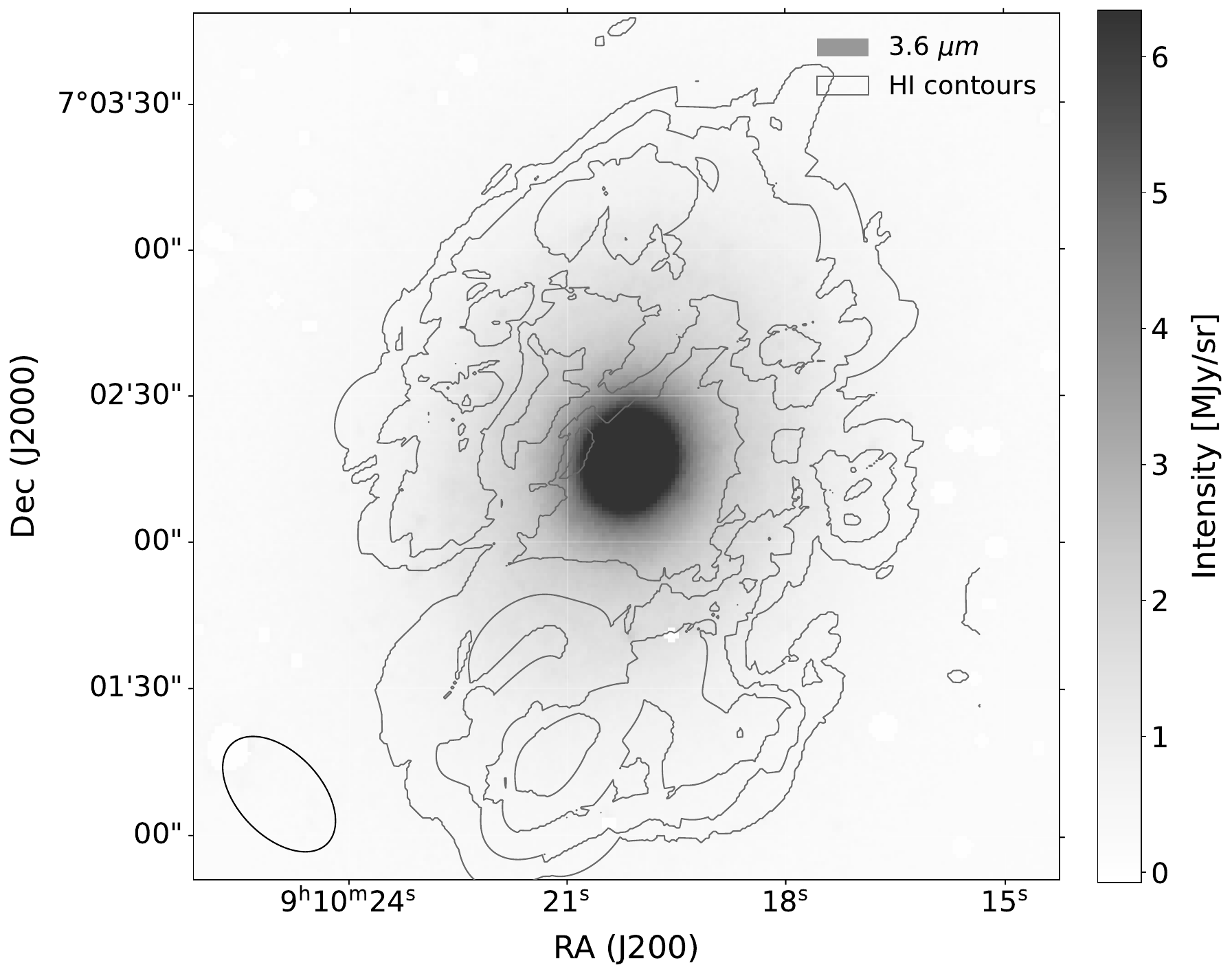}
    \includegraphics[width=0.3\linewidth]{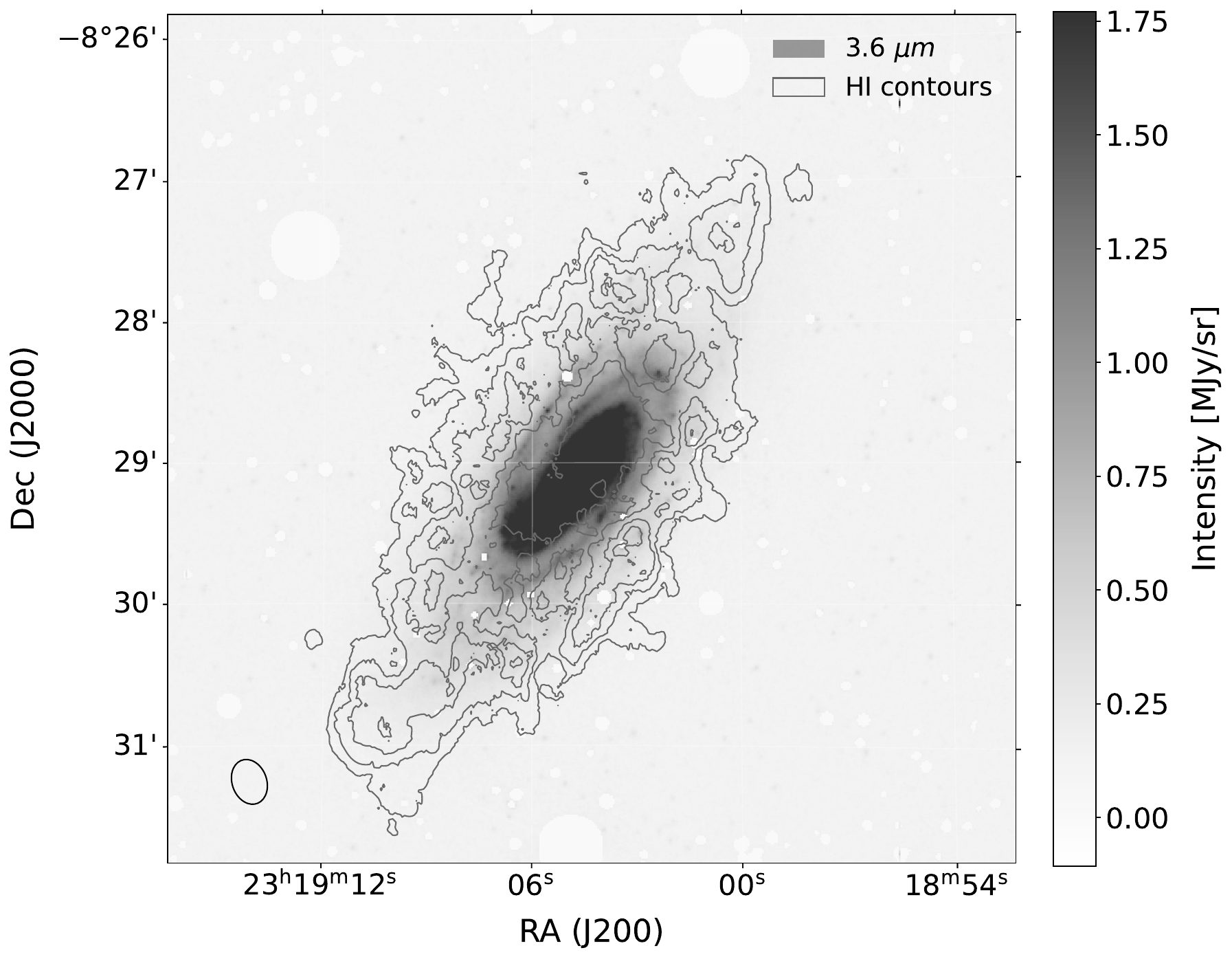}
    \includegraphics[width=0.3\linewidth]{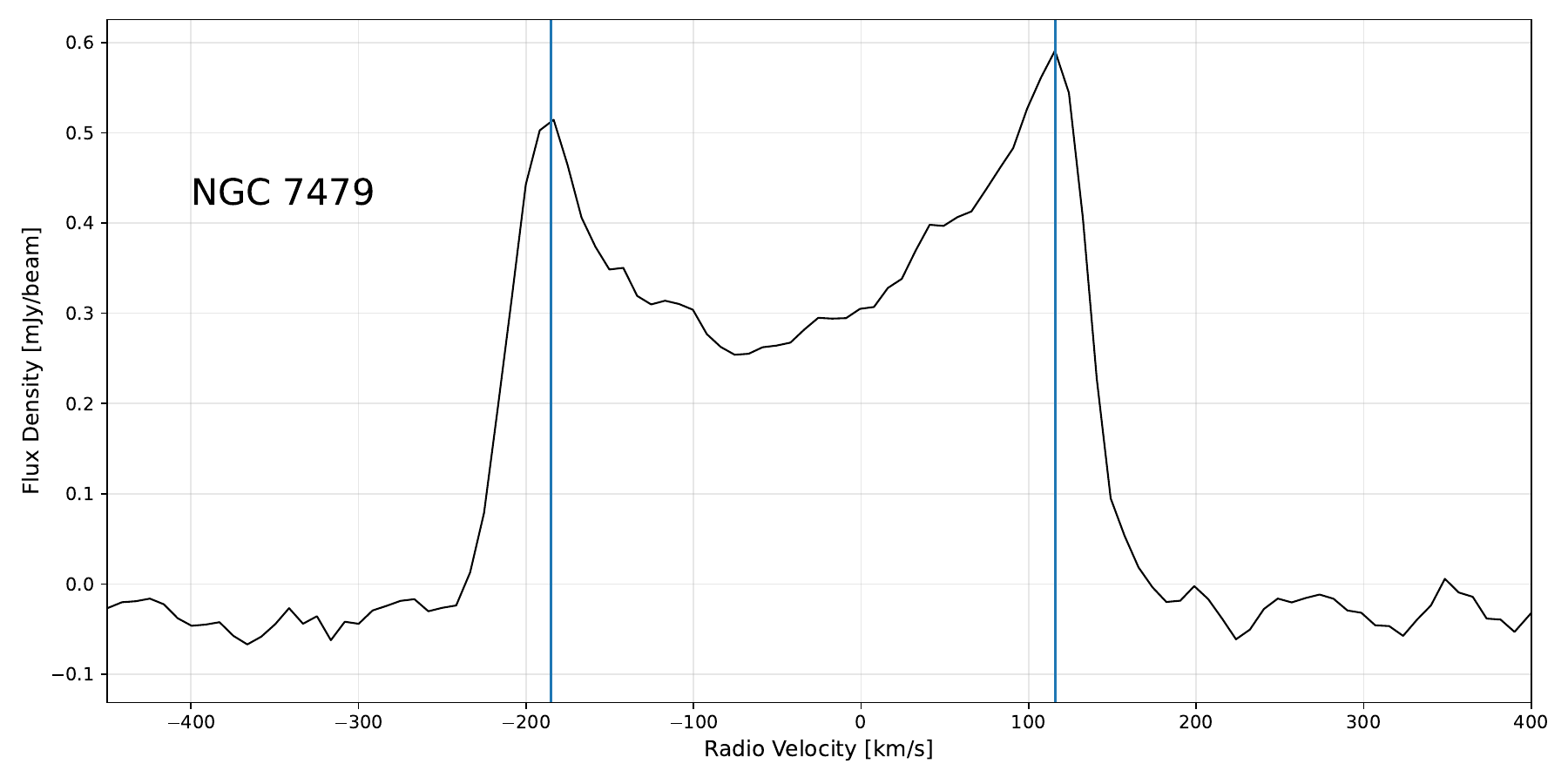}
    \includegraphics[width=0.3\linewidth]{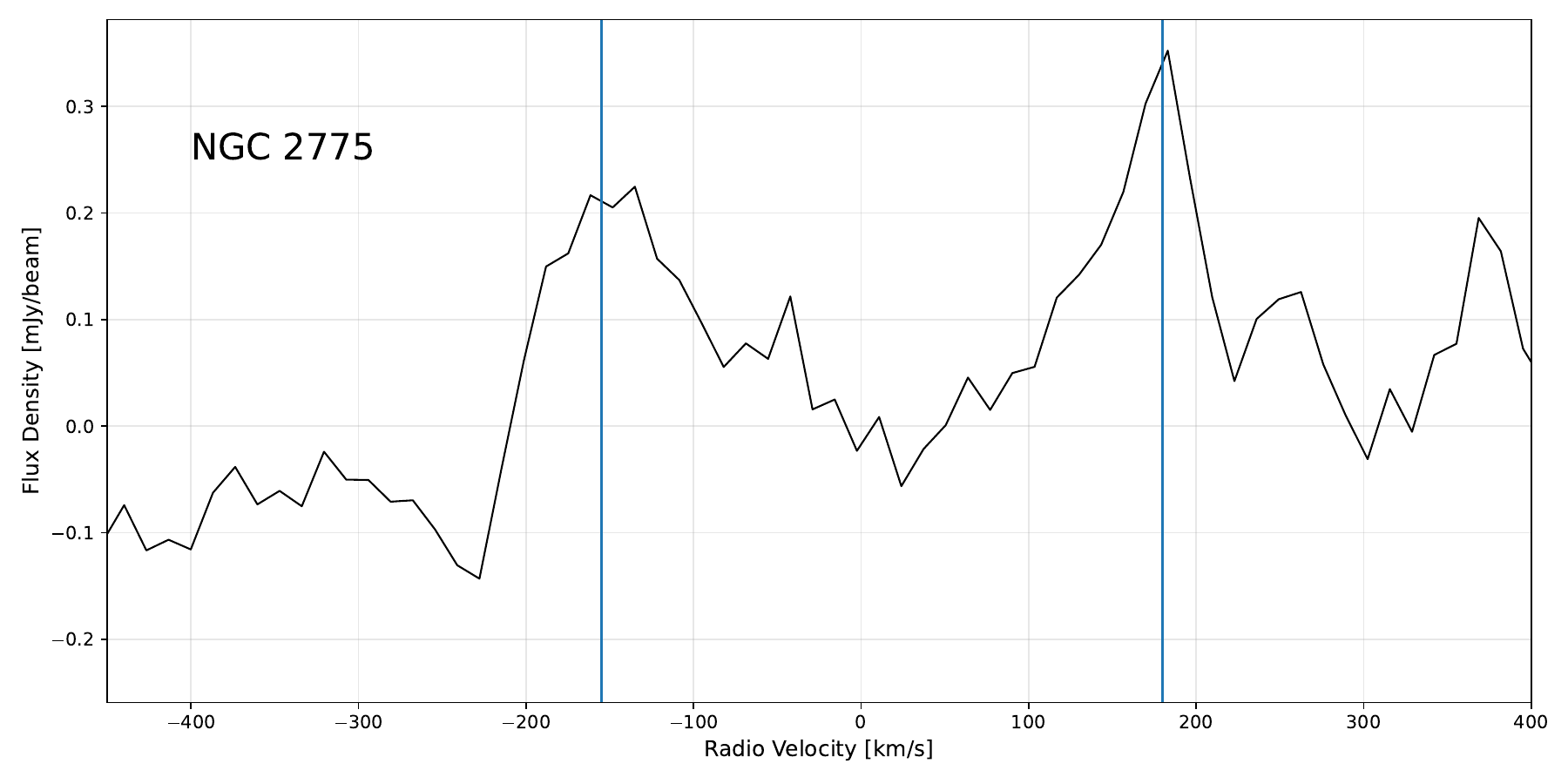}
    \includegraphics[width=0.3\linewidth]{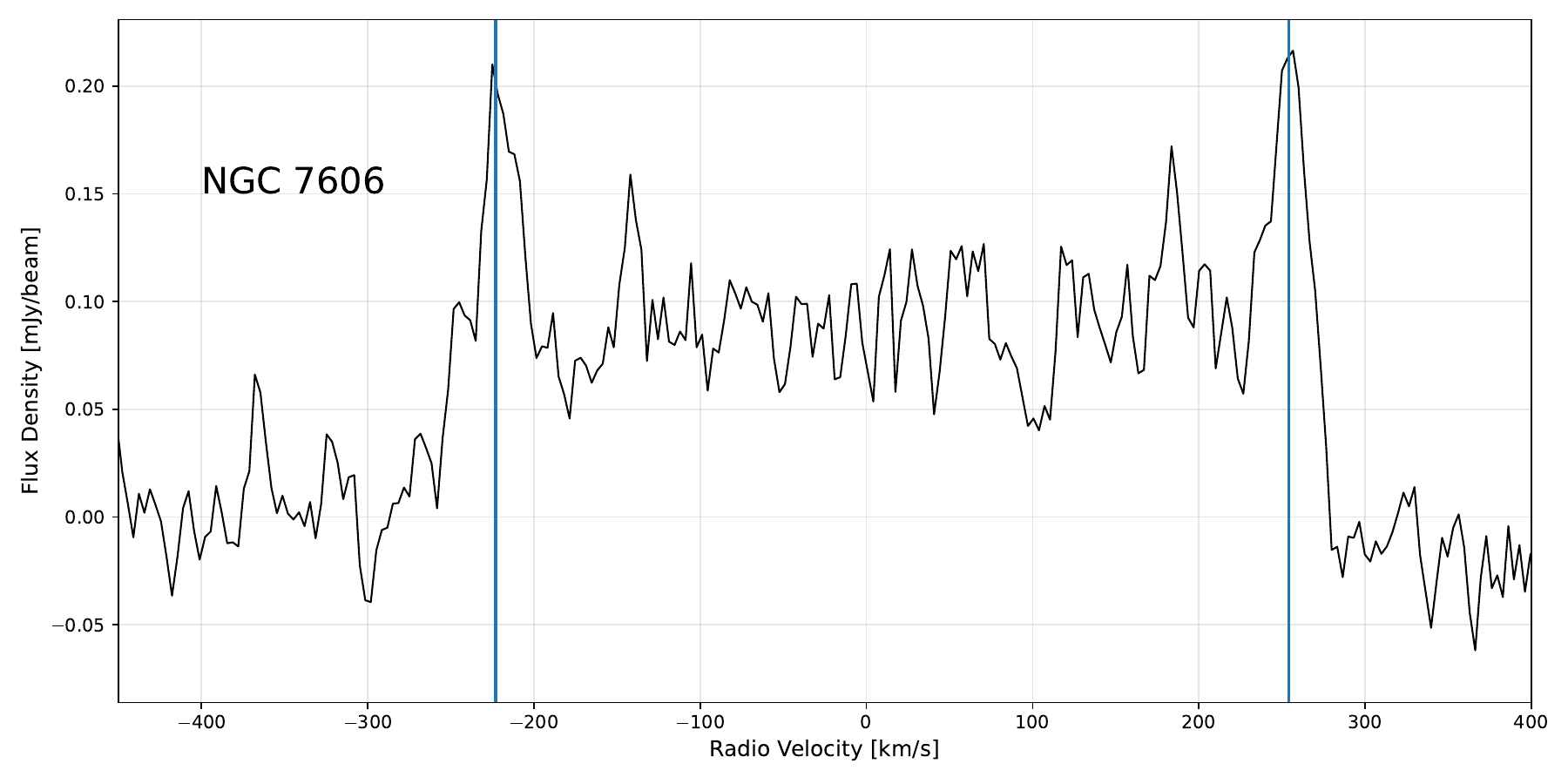}
    \includegraphics[width=0.3\linewidth]{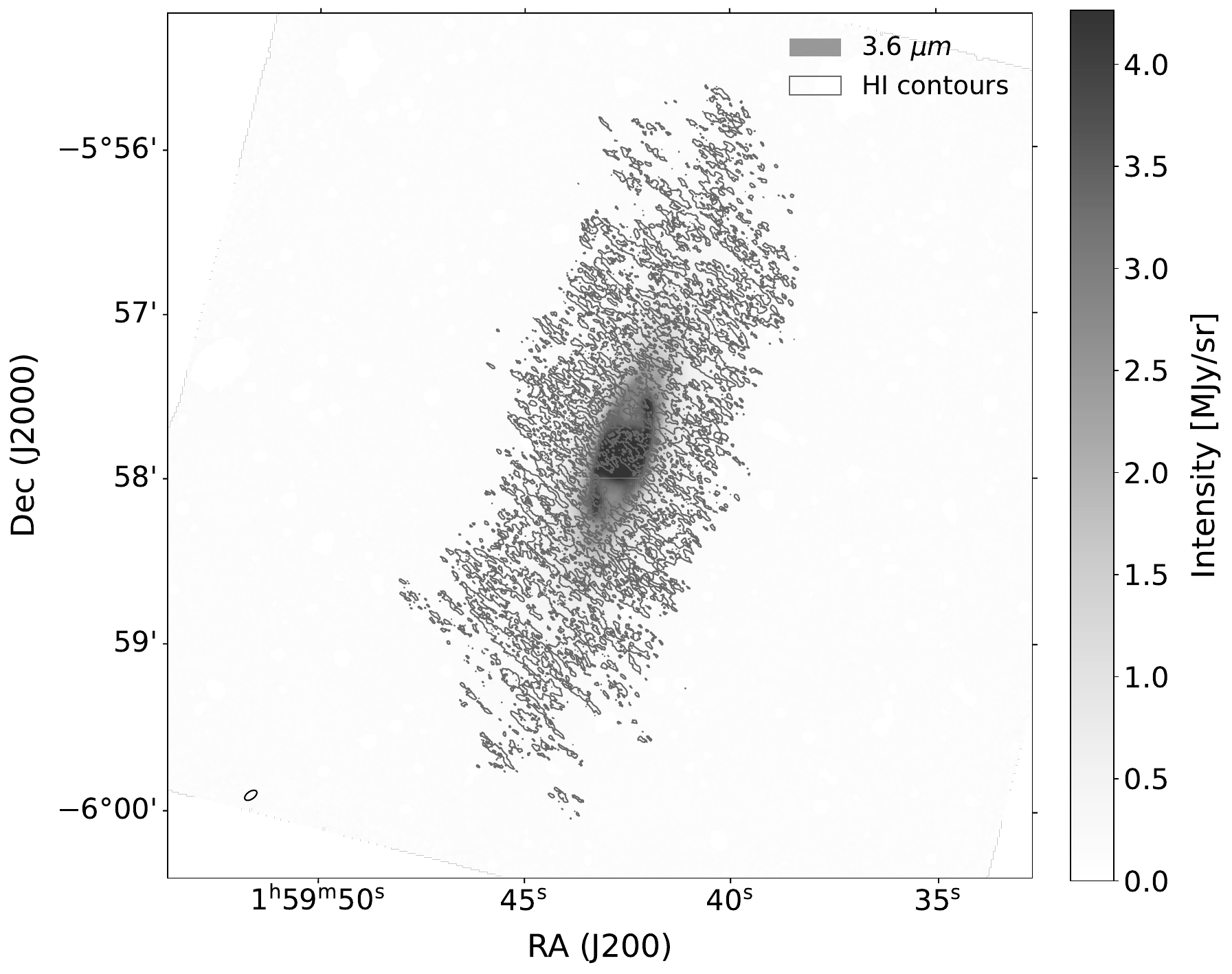}
    \includegraphics[width=0.3\linewidth]{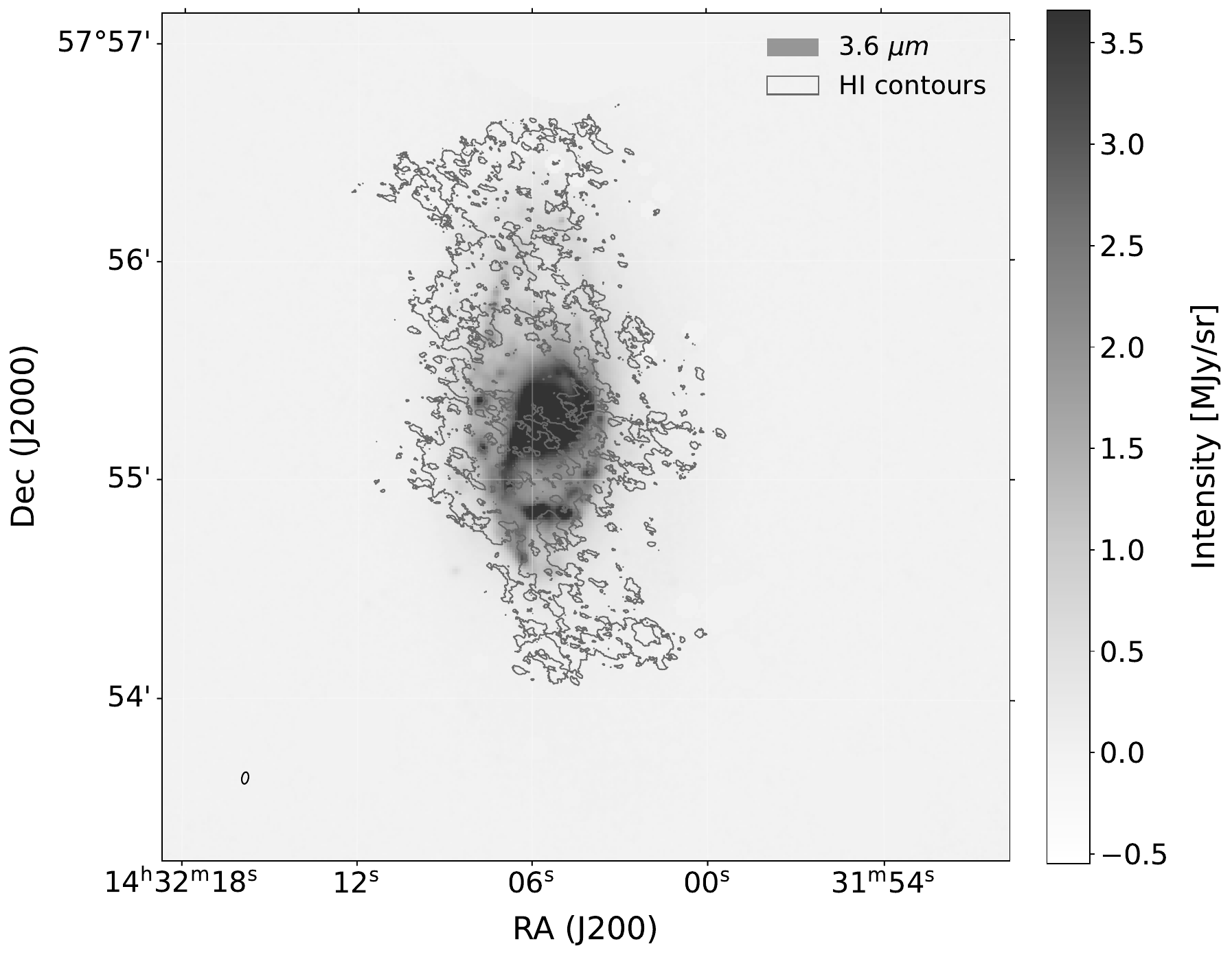}
    \includegraphics[width=0.3\linewidth]{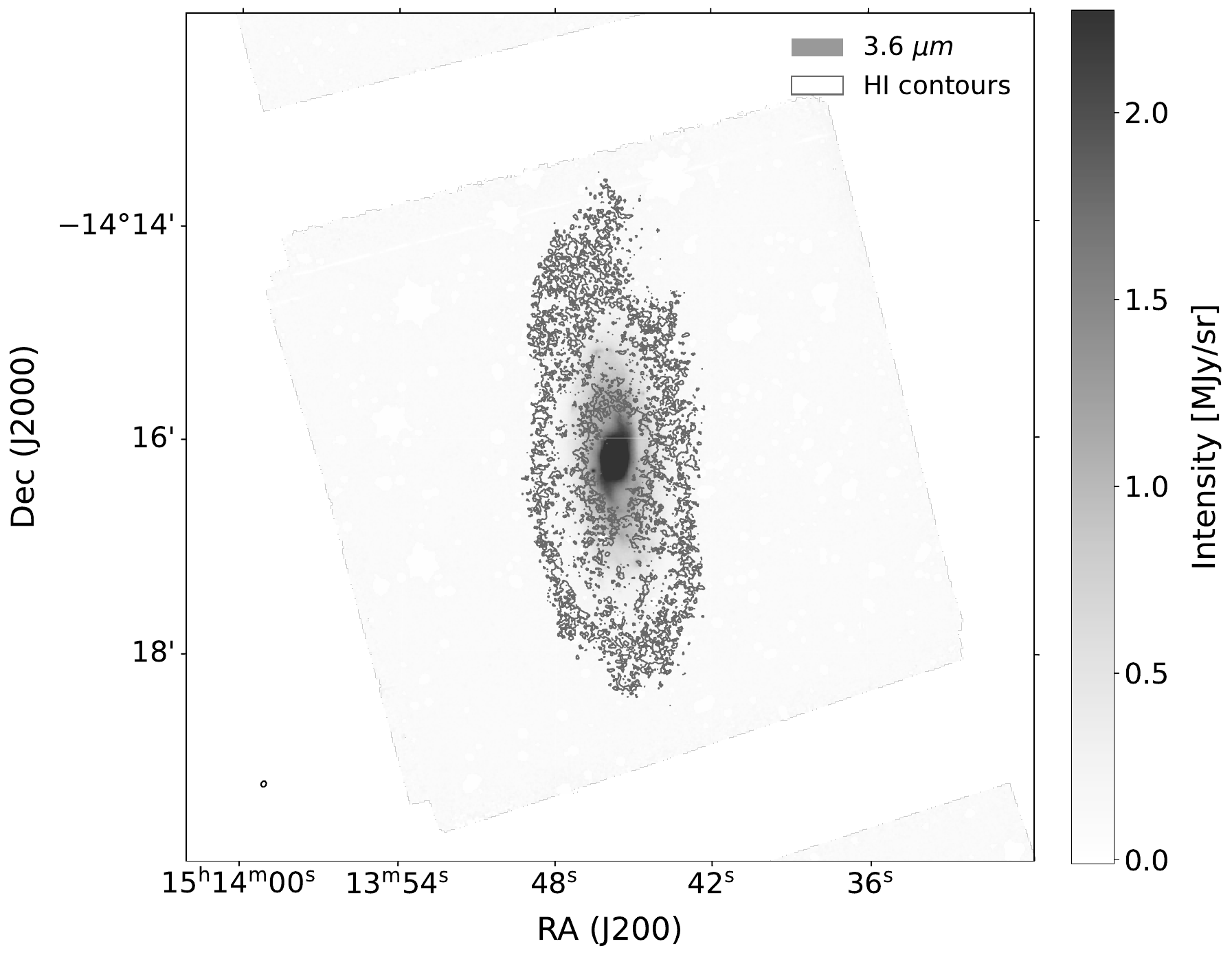}
    \includegraphics[width=0.3\linewidth]{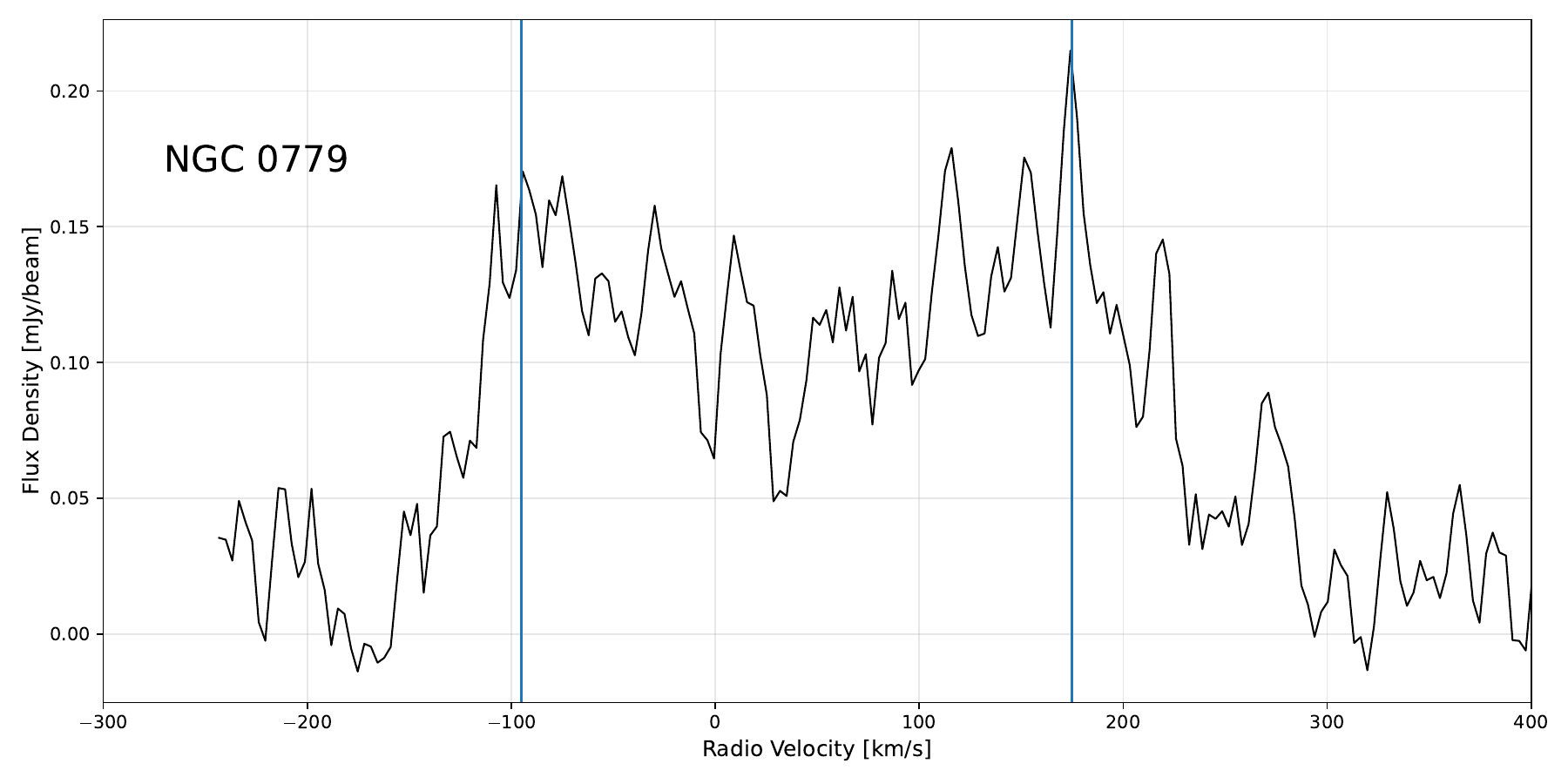}
    \includegraphics[width=0.3\linewidth]{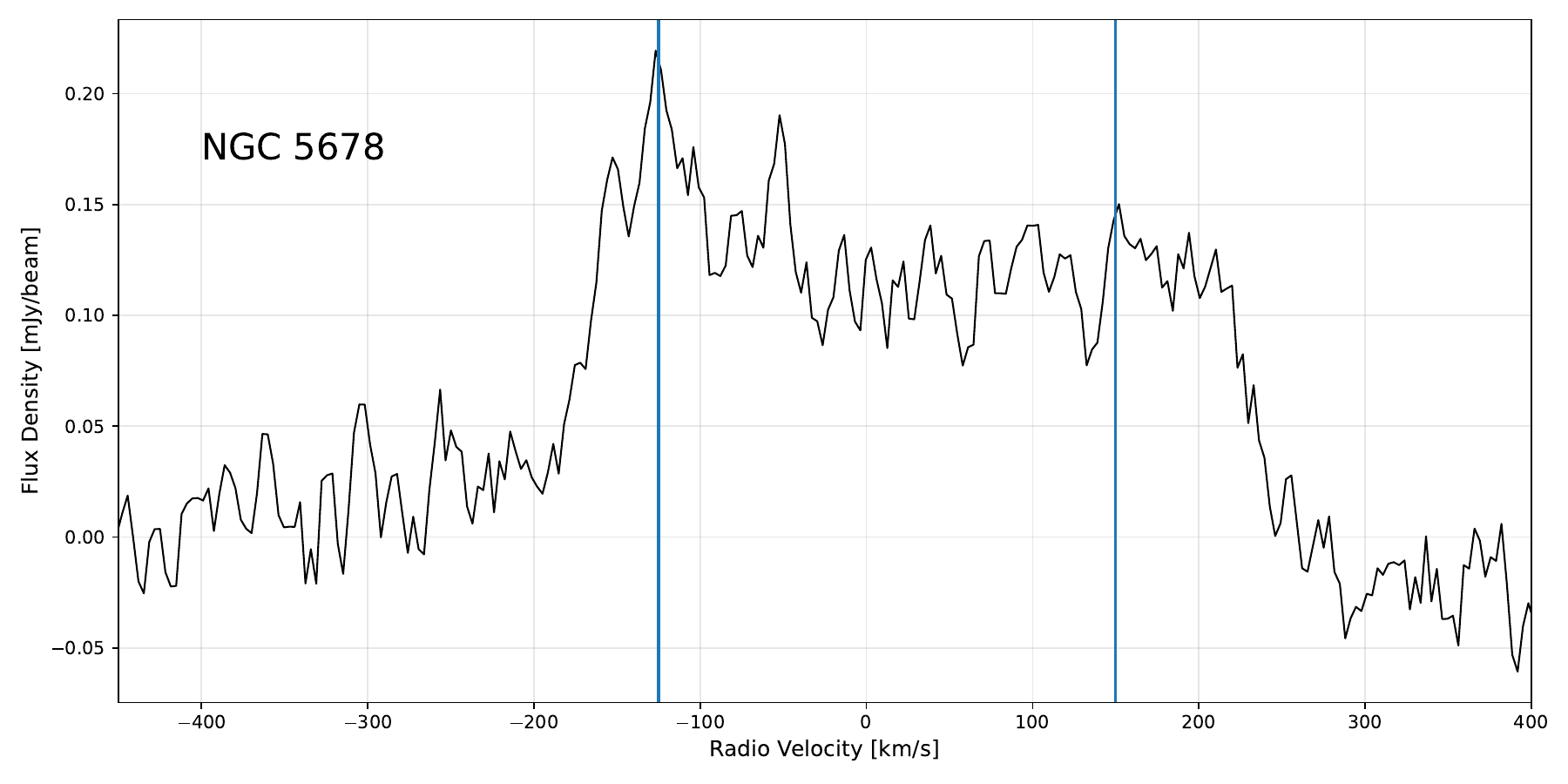}
    \includegraphics[width=0.3\linewidth]{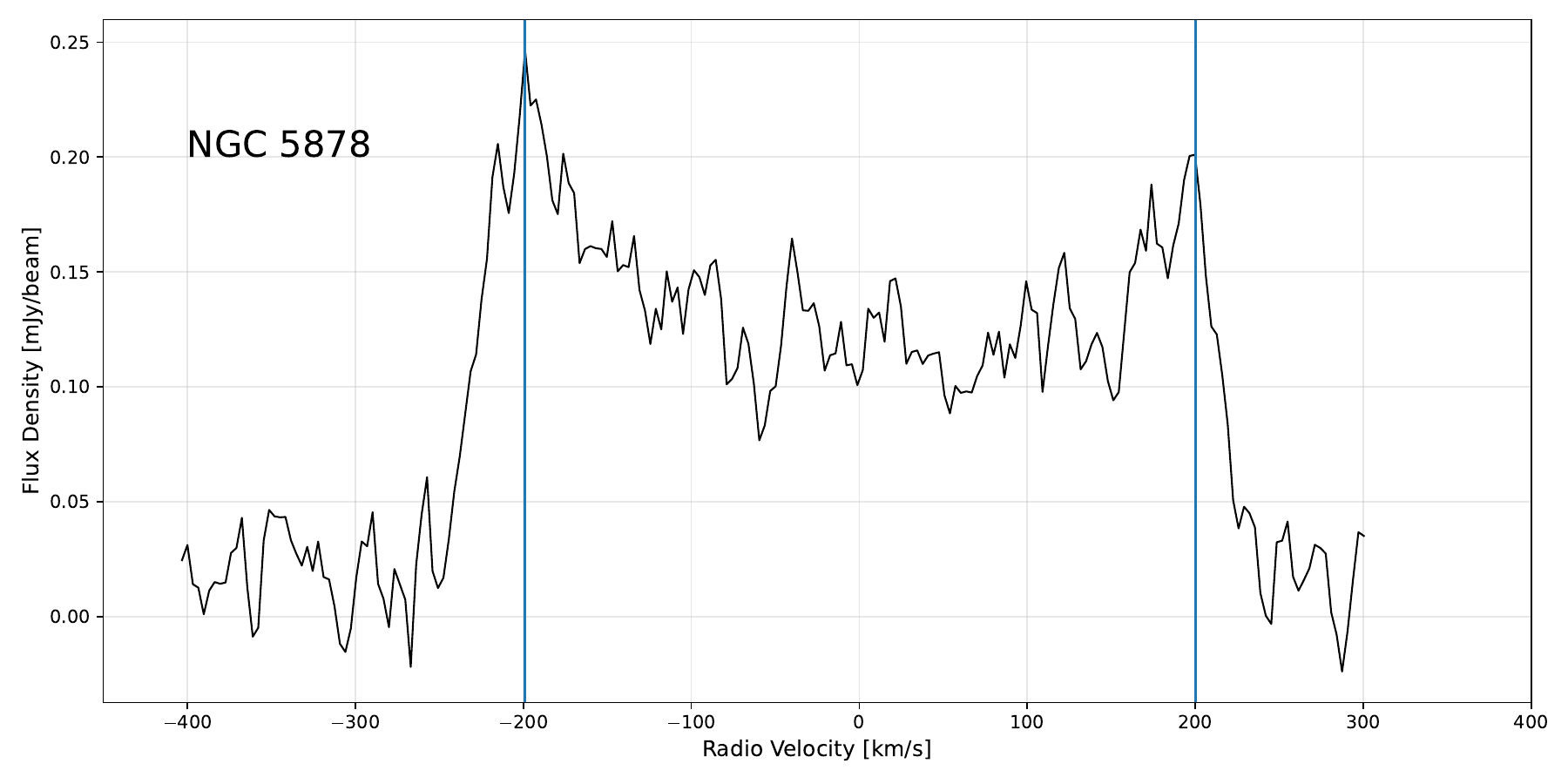}

    \caption{\textit{Rows 1 $\&$ 3:} S$^4$G mid-infrared images \citep{sheth2010} of our six MW analog galaxies with overlaid HI intensity contours, ordered by total observing time. The contours are based on the 0th moment map from \textsc{$^{3D}$Barolo} (see Section \ref{subsec:barolo} for details about the kinematic modeling). Contours are shown at 1, 2, 3, and $5\sigma$ levels for VLA sources, and at 2 and $5\sigma$ levels for GMRT sources.
    \textit{Rows 2 $\&$ 4:} HI line spectra for the observed galaxies, corrected for optical redshifts. 
    While the characteristic double-horn profile (indicated by the blue vertical lines) of a rotating disk is clearly detected in all galaxies, the GMRT data exhibit higher noise levels. See text for details.
    }
    \label{fig:observed}
\end{figure*}

In Figure \ref{fig:observed} we also show the resulting HI integrated intensity maps and line spectra for the galaxies observed with the GMRT.
These sources exhibit RMS noise levels of $8.3-8.8\cdot 10^{-4}$, approximately 20 times higher than those achieved on NGC~7479 and NGC~2775 with the VLA. 
For the GMRT data reduction, we used the \textsc{TCLEAN} task with \textsc{wproject} gridder and \textsc{multiscale} deconvolver for NGC~0779, and \textsc{standard} gridder with \textsc{hogbom} algorithm for NGC~5678 and NGC~5878.

It is worth mentioning that the high spatial resolution achieved by mainly long baselines of the interferometer prevented the detection of large-scale, diffuse emission. This arises from a limited number of short baselines, which creates gaps in the uv-coverage that correspond to extended structures. 
Consequently, despite its superior spectral and spatial resolution, the GMRT data appear to recover only a fraction of the total HI flux, missing a significant portion of the large-scale, diffuse HI emission that dominates the outer disks (see Figure \ref{fig:observed} for comparison with VLA observations).
We therefore acknowledge that the derived HI mass estimates are likely lower limits.

\section{MW ANALOGS FROM TNG50}\label{sec:tng50}

We expand the sample of MW analogs in this work by including simulated MW analogs from the IllustrisTNG suite (The Next Generation; \citealt{pillepich,NelsonTNG}) of cosmological hydrodynamical simulations. The IllustrisTNG represents a state-of-the-art framework for modeling universe evolution from redshift $z=127$ to $z=0$, featuring a comprehensive model of baryonic physics. These simulations incorporate stellar and AGN feedback, magnetic fields, and many \textit{subgrid} astrophysical processes, achieving unprecedented resolution. 
With sophisticated feedback prescriptions and a 50 Mpc-on-a-side volume that affords high resolution ($2160^3$ gas and DM cells, $m_{DM}=8.5\times10^4$), IllustrisTNG-50 (hereafter TNG50; \citealt{Pillepich_2019,Nelson_2019}) resolves galactic scales down to $\sim100~ \mathrm{pc}$, enabling detailed comparisons between simulated and observed galaxy populations.

Similarly to our observational selection, we aim to select late-type galaxies similar to the MW, with similar DM halos. Considering the readily-available properties that the TNG50 database provides for all sub-halos, we based our selection on two main properties: the maximum rotation velocity ($V_{\mathrm{max}}$) and the star formation rate (SFR).
The maximum rotation velocity ($V_{\mathrm{max}}$), which includes the gravitational contribution of all particle types (gas, stars, DM, and black holes), allows for the selection of DM halos with total masses comparable to the MW's \citep{Posti_2019}.  We opted for the SFR criterium as a means to isolate late-type (disk) galaxies, which are characterized by active star formation. 
Based on these considerations, we select TNG50 subhalos at redshift $z=0$ (Snapshot 99 in TNG50) requiring: (i) $1.0 < \mathrm{SFR} < 8.5~M_{\odot}~\mathrm{yr}^{-1}$, consistent with MW values \citep{Elia_2022}; and (ii) $200 < V_{\mathrm{max}} < 315~\mathrm{km}~\mathrm{s}^{-1}$, based on the established range for the MW's circular velocity \citep{reid2016}
Applying these criteria resulted in a sample of 141 simulated MW-like systems. We further eliminate 14 systems with little or no DM particles ($M_{\star}/M_{DM}~>~50~\%$). We attribute these close-to-zero DM halos as potential results of galaxy mergers from which ``chunks" of stellar mass were separated from their parent halos. These TNG50 subhalos, although they comply with our main selection criteria, appear to be environments in which all the DM was removed. We refer to these as systems with no cosmological origin, and we exclude them from our analysis. This result in a final sample of 127 simulated MW analogs from TNG50 suite.

\begin{deluxetable*}{lccccccccccc}
\tabletypesize{\scriptsize}
\tablewidth{0pt} 
\tablecaption{Observational information and inferred properties for our sample of MW analog galaxies \label{tab:sources}}
\tablehead{
\colhead{Object} & \colhead{RA} & \colhead{DEC} &  \colhead{Distance} & \colhead{$\Delta t$} & \colhead{Synt. beam} & \colhead{$\Delta \nu$} & \colhead{PA} & \colhead{Inc.} & \colhead{$M_{HI}$} & \colhead{$M_\star$} & \colhead{$v_{HI}$} \\
\colhead{} & \colhead{[$hh~mm~ss$]} & \colhead{[$dd~mm~ss$]} & \colhead{[Mpc]} & \colhead{[hours]} & \colhead{[$arcsec^{2}$]} & \colhead{[$km~s^{-1}$]} & \colhead{[$^{o}$]} & \colhead{[$^{o}$]} & \multicolumn{2}{c}{[$log_{10}(M/M_{\odot})$]} & \colhead{[$km~s^{-1}$]} \\[-2pt]
\colhead{(1)} & \colhead{(2)} & \colhead{(3)} & \colhead{(4)} & \colhead{(5)} & \colhead{(6)} & \colhead{(7)} & \colhead{(8)} & \colhead{(9)} & \colhead{(10)} & \colhead{(11)} & \colhead{(12)}
}
\startdata
NGC7479 & 23 04 57.6 & +12 19 48 & 33.39 & 6.9 & 20x3 & 9 & 207 & 43.00 & 9.91 & 11.02 & 240 \\
NGC2775 & 09 10 19.2 & +07 02 24 & 17.00 & 4.3 & 25x2 & 13 & 155 & 53.89 & 8.11 & 10.74 & 250 \\
NGC7606 & 23 19 04.8 & -08 28 48 & 31.55 & 2.7 & 15x1 & 3 & 145 & 66.76 & 9.68 & 10.84 & 270 \\
\hline
NGC0779 & 01 59 43.2 & -05 58 12 & 17.68 & 10 & 4x2 & 3 & 160 & 70.66 & 8.83 & 10.15 & 180 
\\
NGC5678 & 14 32 04.8 & +57 55 12 & 29.47 & 10 & 3x0.5 & 3 & 183 & 63.53 & 9.51 & 10.72 & 200 \\
NGC5878 & 15 13 45.6 & -01 25 48 & 30.14 & 10 & 3x0.3 & 4 & 182 & 70.13 & 10.16 & 10.81 & 220 \\
\enddata
\tablecomments{(1) Galaxy identifier; (2) Right ascension and (3) declination coordinates in J200 extracted from NASA/IPAC Extragalactic Database; (4) Galaxy distance extracted from NASA/IPAC Extragalactic Database; (5) Total time on source during our observations; (6) Major and minor synthesized beam axes; (7) Final velocity resolution in the HI data cubes; (8) Position angle and (9) Inclination inferred using the \textsc{$^{3D}$Barolo} algorithm; (10) Total HI mass calculated from the integrated HI flux using Equation~\ref{eq:massHI} \citep{catinella2010}; (11) Total stellar mass derived from 3.6 and 4.5 $\mu$m fluxes using Equation~\ref{eq:massStar} \citep{querejeta2015}; (12) Maximum rotation velocity from the HI-derived RCs (see Figure \ref{fig:RCs}).} 
\end{deluxetable*}

\section{Isolating the DM component in observed and simulated MW analog samples}\label{sec:methods}

The RC of a galaxy is a powerful dynamical tool for deriving its mass distribution, particularly for late-type systems \citep[e.g.][]{Sofue2015}. Decomposing the RC into its baryonic (stellar and gaseous) and DM components provides critical constraints on the properties of the DM halo, including its mass and radial extent \citep{Di_Cintio_2013,DiTeodoro2022, deisídio2024}. In this work, we isolate the DM component in galaxies from both observational and simulation-based perspectives. The observational part involves the kinematic modeling of radio interferometric and mid-infrared data. In parallel, we analyze a wider sample of galaxies from the TNG50 cosmological simulation. 

%
%
\subsection{Rotation curve construction via kinematic modeling of HI data}\label{subsec:barolo}

Building RCs is an effective way of tracing the gravitational potential of spiral galaxies. These curves are typically derived through kinematic modeling, for which the tilted-ring model has provided an effective and widely used framework \citep[e.g.,][]{Rogstad1974}. 
This approach describes a galaxy as a series of concentric rings, each with its own kinematic properties. The main assumption is that the emitting gas is confined to a thin, rotating disk, and that the motion within each ring is dominated by a constant circular velocity, dependent only on the galactocentric radius. 
Traditionally, tilted-ring modeling was applied to two-dimensional (2D) velocity maps derived from observational data. However, a critical limitation of using 2D maps is their susceptibility to beam smearing, an effect that can severely bias the derived RCs, particularly in a galaxy's inner regions \citep{BosmaPhD,Kamphuis,Biswas_2023}. To overcome these limitations, we employ the \textsc{$^{3D}$Barolo} software \citep{diteodoro20153dbarolonew3dalgorithm}, which constructs a fully 3D kinematic model of the galaxy directly from the HI data cube. This approach fits tilted rings in 3 dimensions, incorporating iterative corrections for inclination, position angle, and other parameters. By explicitly accounting for instrumental effects during the model convolution step \citep[see][for further details]{ diteodoro20153dbarolonew3dalgorithm}, \textsc{$^{3D}$Barolo} effectively mitigates biases from beam smearing and projection effects.

\begin{figure}
    \centering
    \includegraphics[width=0.4\textwidth]{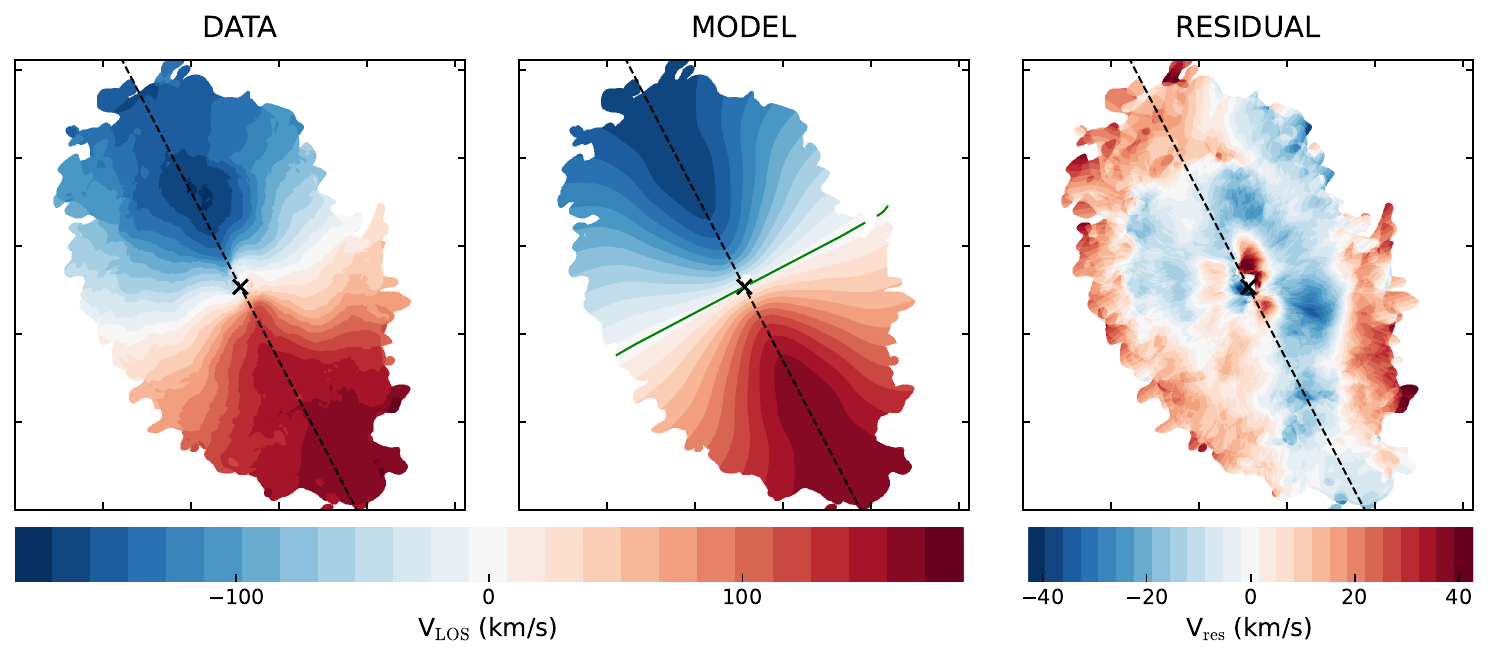}\label{fig:ngc7479}
    \includegraphics[width=0.4\textwidth]{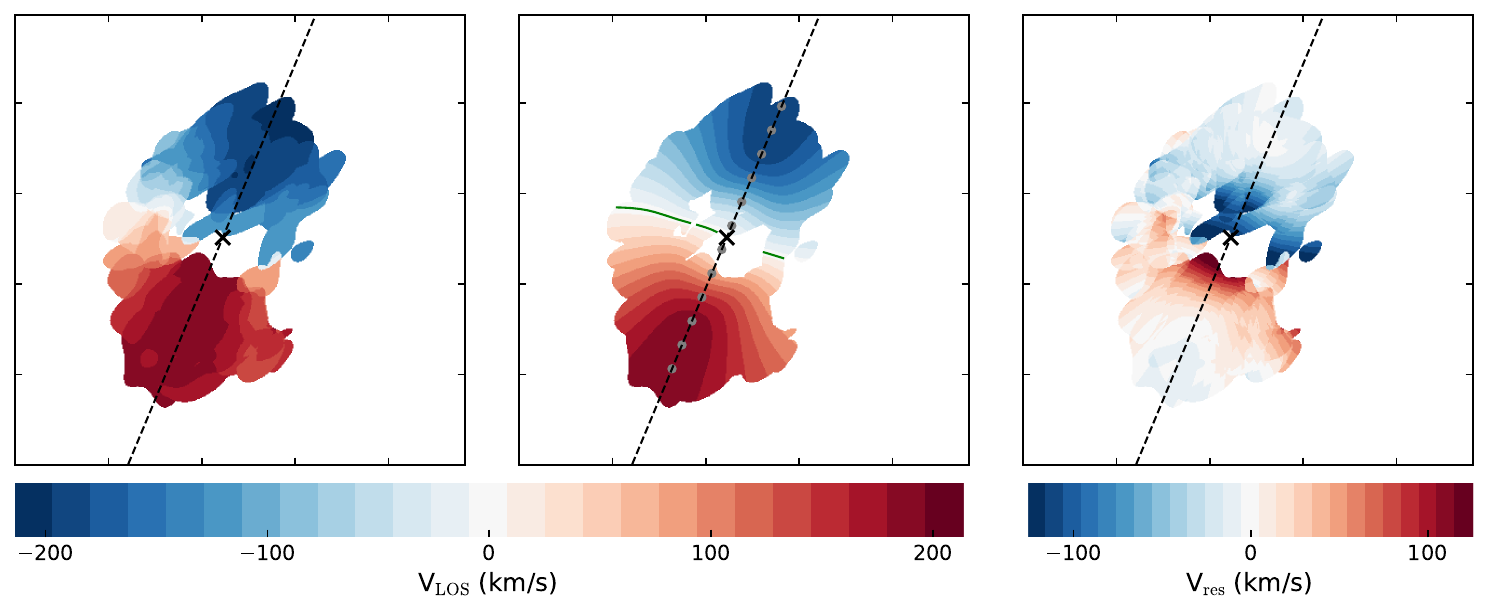}\label{fig:ngc2775}
    \includegraphics[width=0.4\textwidth]{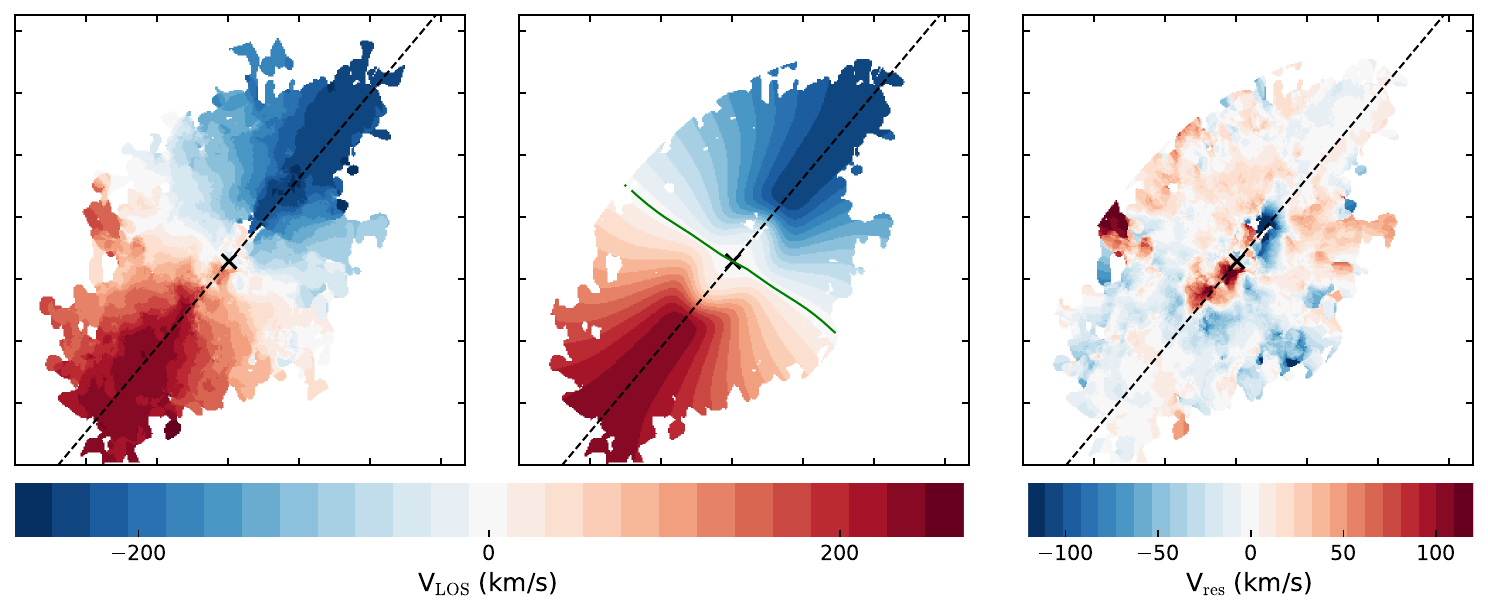}\label{fig:ngc7606}
    \includegraphics[width=0.4\textwidth]{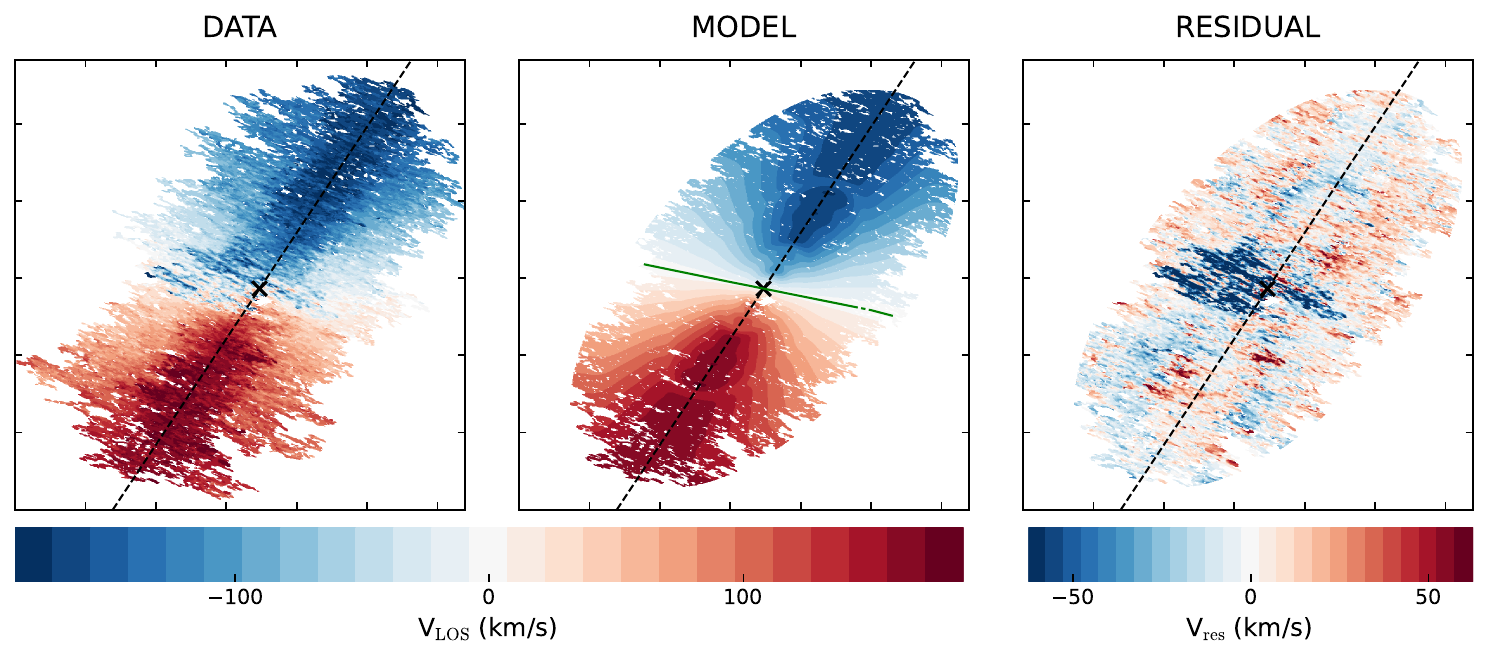}\label{fig:ngc0779}
    \includegraphics[width=0.4\textwidth]{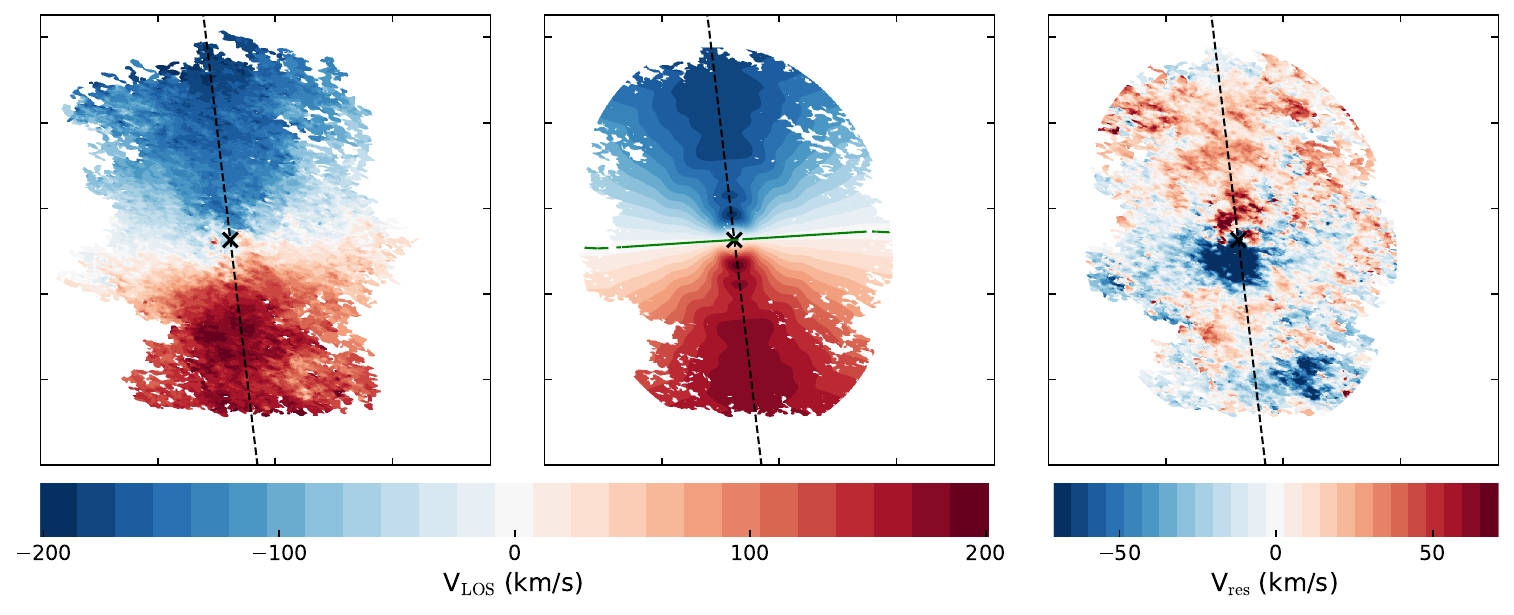}\label{fig:ngc5678}
    \includegraphics[width=0.4\textwidth]{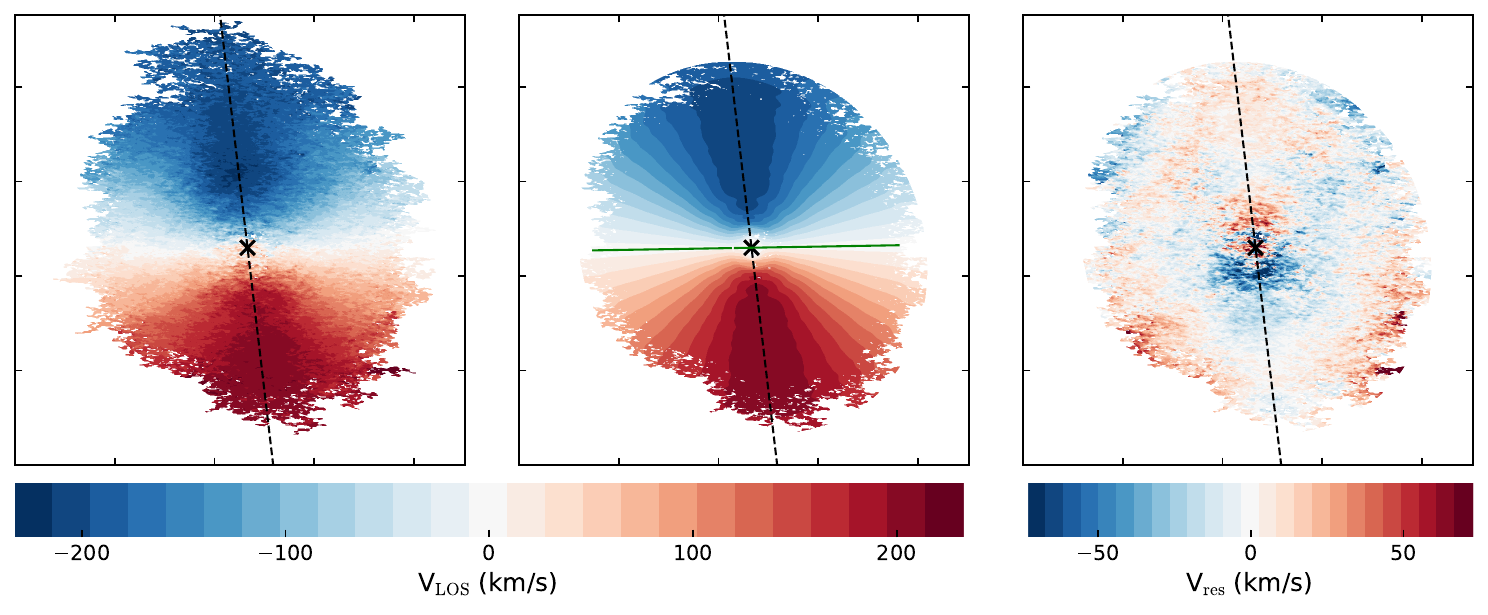}\label{fig:ngc5878}
    \caption{Velocity maps (1st moment) for the observed galaxies: (a) NGC 7479, (b) NGC 2775, (c) NGC 7606, (d) NGC 0779, (e) NGC 5678, and (f) NGC 5878. First column: HI velocity field derived from the radio data cubes. Second column: 2D projection of the 3D kinematic model generated by \textsc{$^{3D}$Barolo}. Third column: Residuals (data $-$ model). The dashed black curve traces the kinematic major axis for each galaxy, as determined by the best-fit inclination and position angle parameters from the \textsc{$^{3D}$Barolo} modeling. The solid green line indicates where the kinematic transition occurs between the receding (redshifted) and approaching (blueshifted) sides.}
    \label{fig:velocities}
\end{figure}

We performed an initial run to estimate the galaxy center coordinates, inclination, and position angle starting guesses. Input distances and initial guesses for maximum rotation velocities were adopted from the \textit{S$^4$G} catalog \citep{sheth2010}, while beam properties (major/minor axes, and position angle) were extracted directly from the HI imaging. The radial sampling (number of radii or radii separation between rings) was dynamically adjusted for each galaxy based on beam size, with multiple tests conducted to optimize the RC fit and minimize residuals. 
\textsc{$^{3D}$Barolo} uses a  mask to identify the regions in the cube where galaxy emission is; to generate this mask we used the built-in \textsc{Search} mode algorithm. 
Figure~\ref{fig:velocities} presents the derived kinematic maps: the line-of-sight velocity (1st moment) and velocity dispersion (2nd moment) for each observed galaxy. Both the receding and approaching sides of the galaxies were fitted. The resulting RCs (shown in Figure X) arise from averaging the final velocity values from both sides at equal radial distances from the center.

%
%

\subsection{Mass decomposition}\label{subsec:mass}

To robustly isolate the DM component, we account for baryonic mass distributions by directly measuring gas and stellar contributions.

\subsubsection{Gas component}

In spiral galaxies, HI typically constitutes $\sim70–80\%$ of the total gas mass \citep{Saintonge_2022}. To account for the contribution of Helium and heavier elements, we apply a correction factor of 1.36 to the derived HI mass \citep{Chowdhury_2022}.  Given that molecular gas ($\mathrm{H_2}$) is not a dominant mass component in the outer disks of spiral galaxies we omit its contribution from our mass model (see \citealt{mancera2022} for $\mathrm{H_2}$ and CO kinematic modeling). To calculate the HI masses we adopt the mass relation from \cite{catinella2010}:

\begin{equation}
    M_{HI} = \frac{2.356 \times 10^{5}}{1+z} \left[ \frac{d_{L}(z)}{Mpc} \right]^{2} \left[ \frac{S_{HI}}{Jy~km~s^{-1}} \right]
\label{eq:massHI}
\end{equation}
    
\noindent where $d_{L}(z)$ is the luminosity distance at redshift z and $S_{HI}$ is the integrated flux. The integrated flux is obtained by summing over all source-detected pixels and velocity channels in the data cube:

\begin{equation}
    S_{HI} = \sum{~F_{HI}~\Delta \nu}
\end{equation}

\noindent where $F_{HI}$ is the HI flux measured in each channel and $\Delta\nu$ is the HI data cube bandwidth. The spatial distribution of this integrated flux is shown in the 0th moment maps presented in Figure~\ref{fig:observed}. The derived HI masses for all observed sources in our sample are presented in Table~\ref{tab:sources}.

The \textsc{$^{3D}$Barolo} algorithm calculates the surface mass density of the cold atomic gas directly from the data cube, enabling a direct derivation of the gravitational potential. This is achieved by summing the HI signal within concentric tilted rings and applying corrections for the galaxy's inclination and position angle.

\subsubsection{Stellar component}

Following the same methodology applied to the gas decomposition, we sum the mid-infrared flux within concentric annuli matching those derived by \textsc{$^{3D}$Barolo} for each HI data cube. This ensures consistent radial sampling when comparing gas and stellar distributions. 
For the stellar mass conversion, we apply the empirical relation from \citet{querejeta2015}: 

\begin{equation}
    M_{\star} = 10^{8.35}~\left[\frac{F_{3.6 \mu m}}{Jy}
\right]^{1.85} \left[ \frac{F_{4.5 \mu m}}{Jy} \right]^{-0.85}  \left[ \frac{D}{Mpc} \right]^{2}
\label{eq:massStar}
\end{equation}

\noindent where $F_{3.6\mu m}$ and $F_{4.5\mu m}$ are the $3.6$ and $4.5\mu m$ band fluxes and $D$ is the galaxy distance. The derived stellar masses for all observed sources in our sample are presented in Table~\ref{tab:sources}.

\subsubsection{Dark matter halo profile}\label{subsec:dmhaloprofile}

Although debates persist regarding a diversity of DM density profiles in late-type galaxies \citep[e.g.,][]{Oman_2015}, the NFW profile remains one of the most widely supported models for reproducing the circular velocities of disc systems \citep{Salucci_2007,Di_Cintio_2013,IoccoPato_2015,PatoIocco_2015}. We assume an NFW profile and determine a set of virial mass ($M_{200}$) and concentration ($c$) parameters for each galaxy, following the methodology detailed in \citet{deisídio2024}, briefly described here.

Considering that the DM halo is gravitationally coupled with stellar and gaseous mass distributions, the inferred halo parameters directly depend on the baryonic kinematics. To estimate $M_{200}$ and $c$ for the RC of each galaxy in our sample, we perform a Markov Chain Monte Carlo (MCMC) analysis - implemented with the Python package \texttt{emcee} \citep{MCMC2013}. For each galaxy, we use 50 walkers and run 30,000 iterations, discarding the first $25\%$ of each chain as burn-in. In addition to this approach from \cite{deisídio2024}, we constrain the parameter space of $M_{200}$ and $c$, by adopting the concentration-mass relation derived by \citet{Dutton_2014}:

\begin{equation}
\log_{10}(c) = 0.905 - 0.101 \log_{10}\left(\frac{M_{200}}{10^{12}~h^{-1}~M_{\odot}}\right).
\end{equation}
 
We set flat priors for the halo mass over the range $9<~log(M_{200}/M_{\odot})<13$ and for the concentration parameter over $1<c<50$.
The resulting best-fit parameters and their uncertainties from our MCMC analysis are reported in Table~\ref{tab:MCMC}.

\begin{deluxetable}{lcccc}
\centering
\tabletypesize{\footnotesize}
\tablewidth{0pt}
\tablecaption{Best-fit halo parameters and LDMD from the MCMC analysis
\label{tab:MCMC}}
\tablehead{
\colhead{Object} & \colhead{$log_{10}(\frac{M_{200}}{M_{\odot}})_{-\sigma}^{+\sigma}$} & \colhead{$c_{-\sigma}^{+\sigma}$} & \colhead{$\frac{\rho_{DM}}{GeV~cm^{-3}}$} & \colhead{\textbf{$\chi^{2}/\nu$}}
\\
\colhead{(1)} & \colhead{(2)} & \colhead{(3)} & \colhead{(4)} & \colhead{(5)}\\
}
\startdata
NGC 7479 & $12.12_{-0.09}^{+0.11}$ & $8.82_{-1.29}^{+1.45}$ & $0.173$ & 0.5173 \\
NGC 2775 & $12.21_{-0.24}^{+0.35}$ & $24.58_{-6.50}^{+7.49}$ & $1.770$ & 8.9909 \\
NGC 7606 & $12.30_{-0.09}^{+0.11}$ & $17.62_{-2.71}^{+3.13}$ & $0.308$ & 0.6678 \\
NGC 0779 & $11.64_{-0.12}^{+0.15}$ & $24.39_{-4.40}^{+5.12}$ & $1.230$ & 0.6880 \\
NGC 5678 & $11.90_{-0.16}^{+0.21}$ & $17.43_{-3.63}^{+4.17}$ & $0.197$ & 0.5387 \\
NGC 5878 & $11.47_{-0.04}^{+0.04}$ & $26.24_{-3.15}^{+3.57}$ & $0.233$ & 0.3638
\enddata
\tablecomments{(1) Galaxy identifier; (2) Halo mass $M_{200}$ and (3) concentration parameter $c$, showing their best-fits values with respective lower and upper errors from the MCMC analysis; (4) DM density at the solar equivalent position in each MW analog ($R_{\odot}=1.8\times R_{eff}$, see Section \ref{subsec:local} for details); (5) Reduced $\chi^{2}$ statistic.}
\end{deluxetable}

\subsection{Rotation curve decomposition}\label{subsec:rc}

To derive the circular velocity contribution from each baryonic component, we calculate the gravitational potential generated by a razor-thin axisymmetric disk with a given surface mass density $\Sigma_{baryon(R)}$:

\begin{equation}
    \Phi_{baryon} = -4~G~\int_0^{r_{max}} R'~\Sigma_{baryon(R')}\frac{K\left( \frac{4RR'}{(R + R')^{2}} \right)}{|R+R'|} dR'
\end{equation}

\noindent where $G$ is the gravitational constant, $\Sigma_{\rm baryon}(R')$ is the surface density at radius $R'$, and $K(m)$ is the complete elliptic integral of the first kind, defined as:

\begin{equation}
    K(m) = \int_{0}^{\frac{\pi}{2}} \frac{d\theta}{\sqrt{1- m sen^{2}\theta}}
\end{equation}

The total circular velocity model, $V_{mod(R)}$, is constructed from the quadratic sum of the individual contributions from the stellar, gaseous, and DM halo components:

\begin{equation}
    V_{mod} = \sqrt{V_{\star}|V_{\star}|~+~V_{gas}|V_{gas}|~+~V_{DM}|V_{DM}|}
\label{eq:vmod}
\end{equation}

\begin{figure*}
    \centering
    \includegraphics[width=0.32\linewidth]{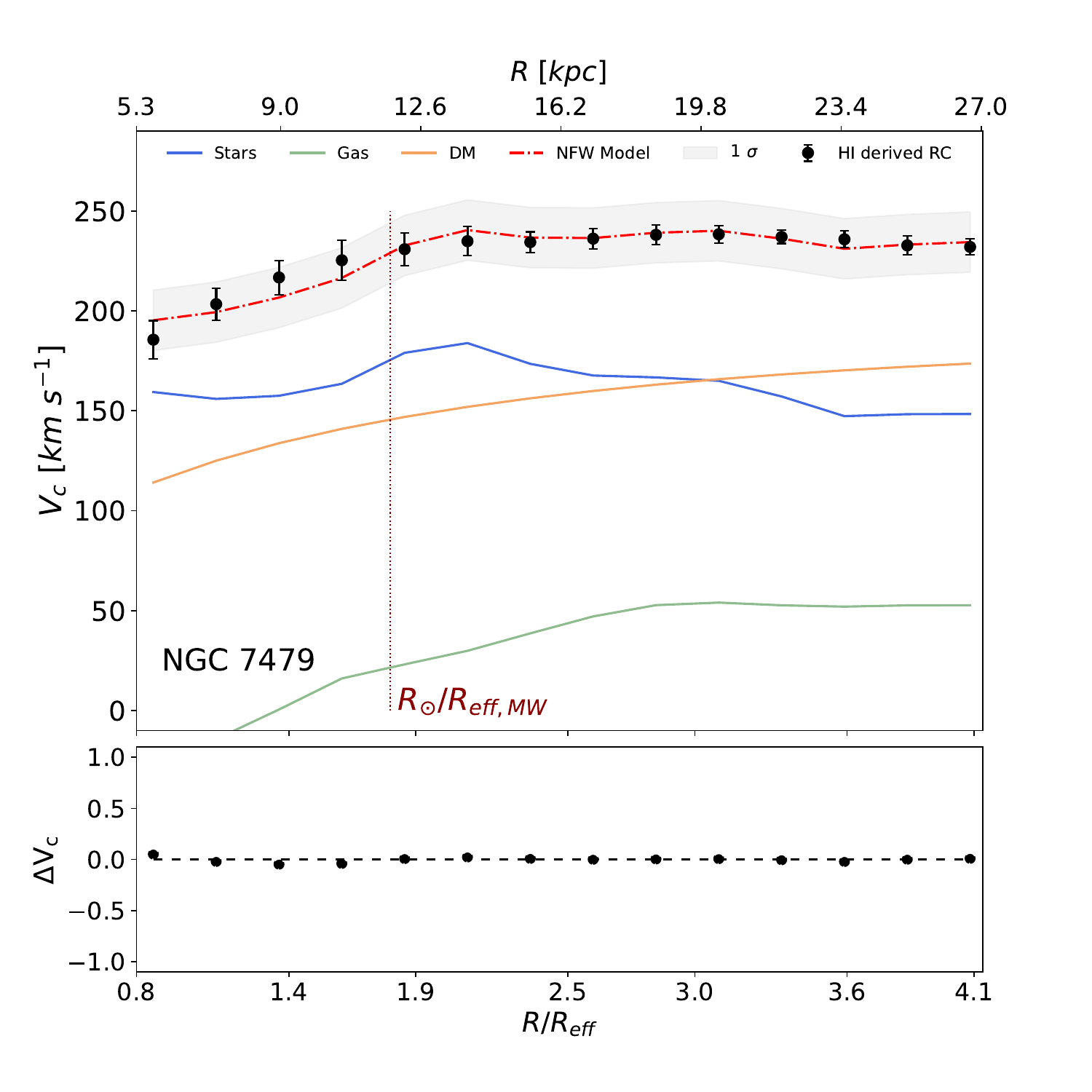}
    \includegraphics[width=0.32\linewidth]{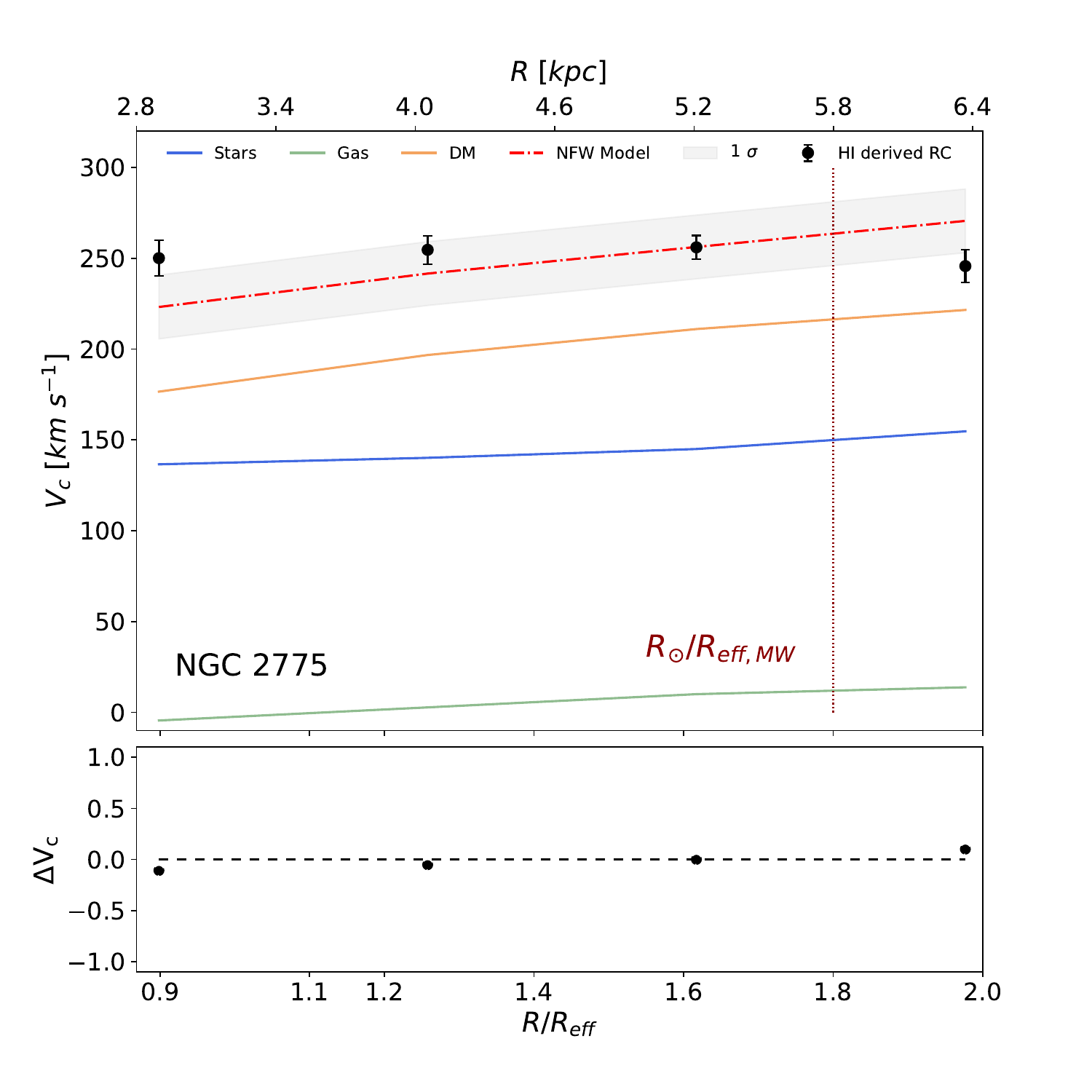}
    \includegraphics[width=0.32\linewidth]{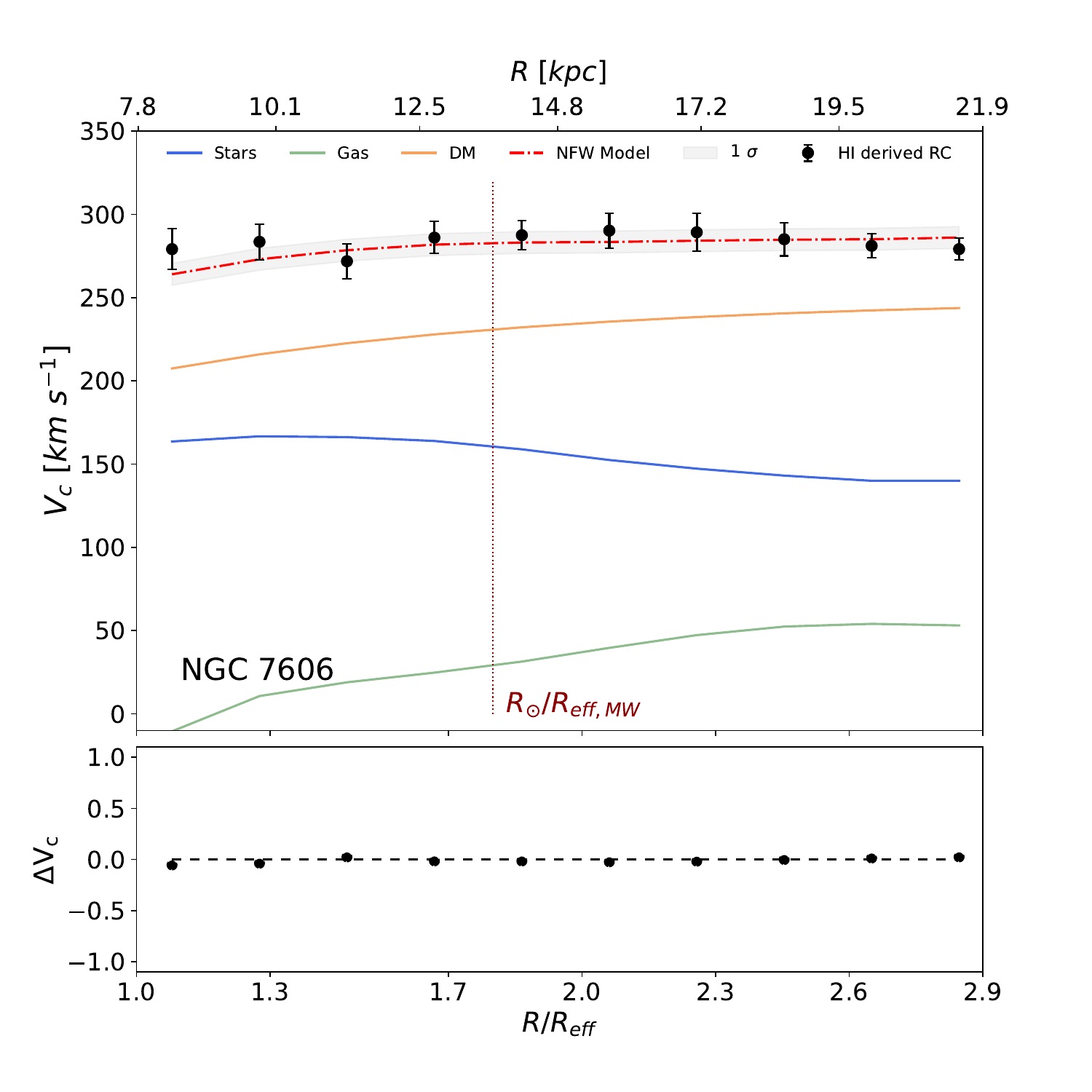}
    \includegraphics[width=0.23\linewidth]{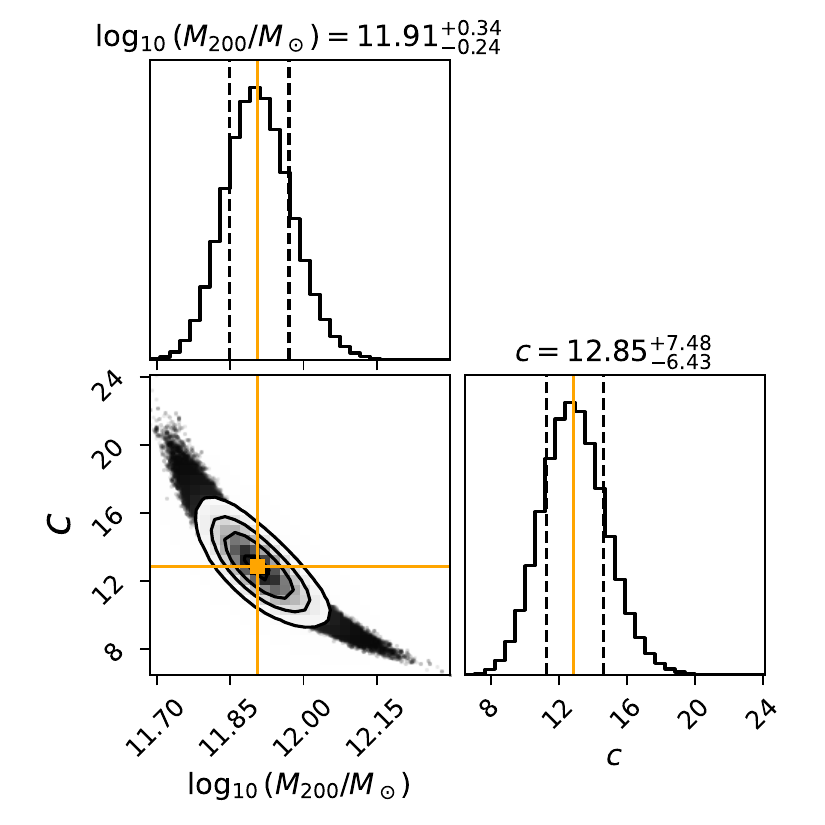}
    \includegraphics[width=0.23\linewidth]{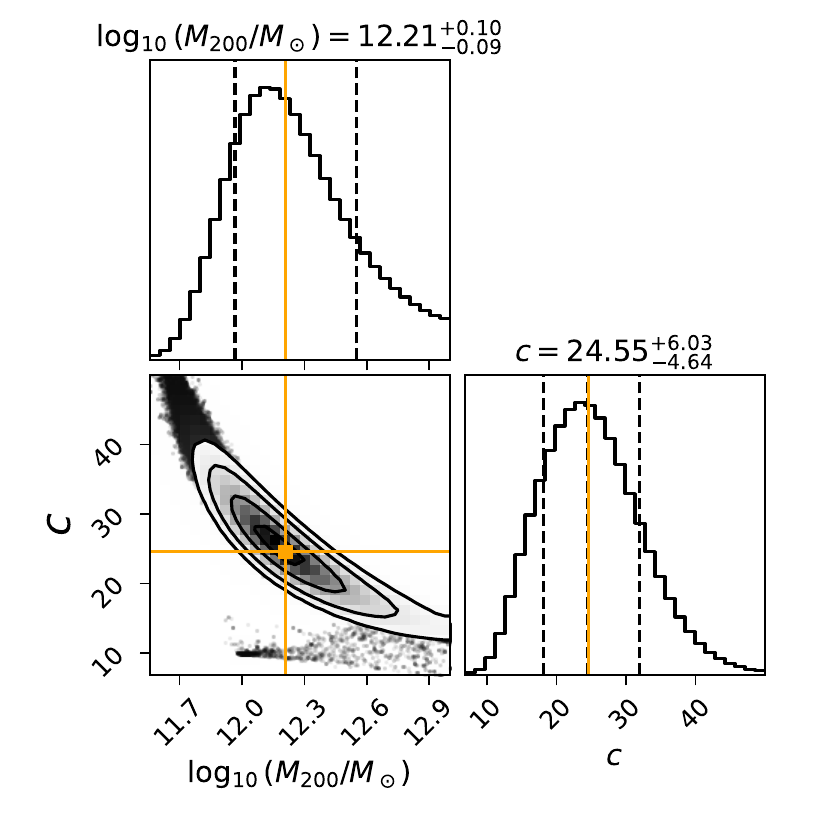}
    \includegraphics[width=0.23\linewidth]{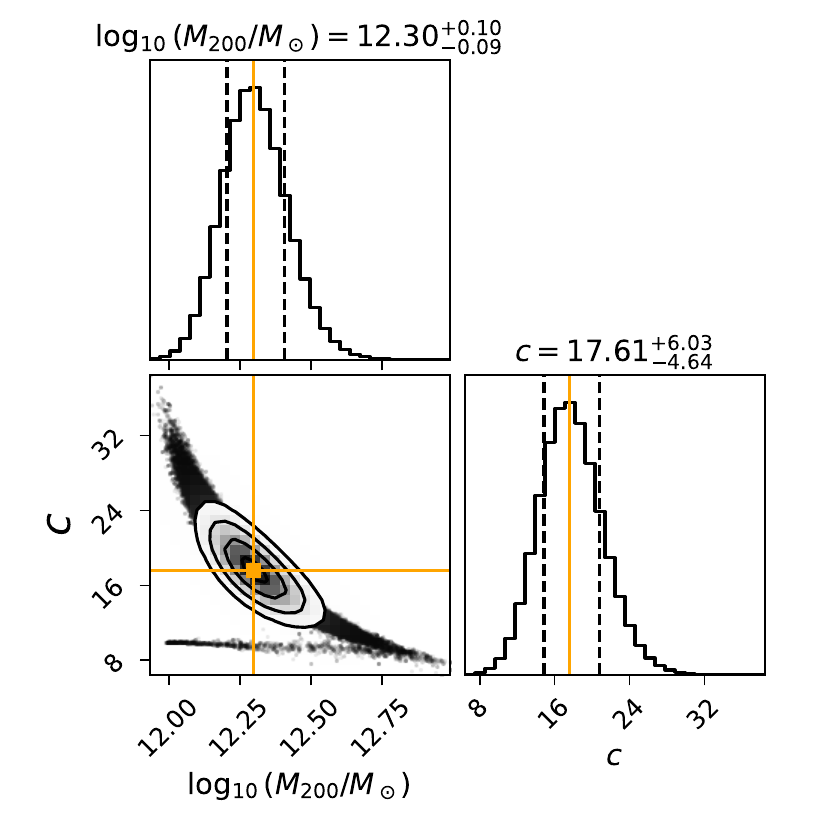}
    \includegraphics[width=0.32\linewidth]{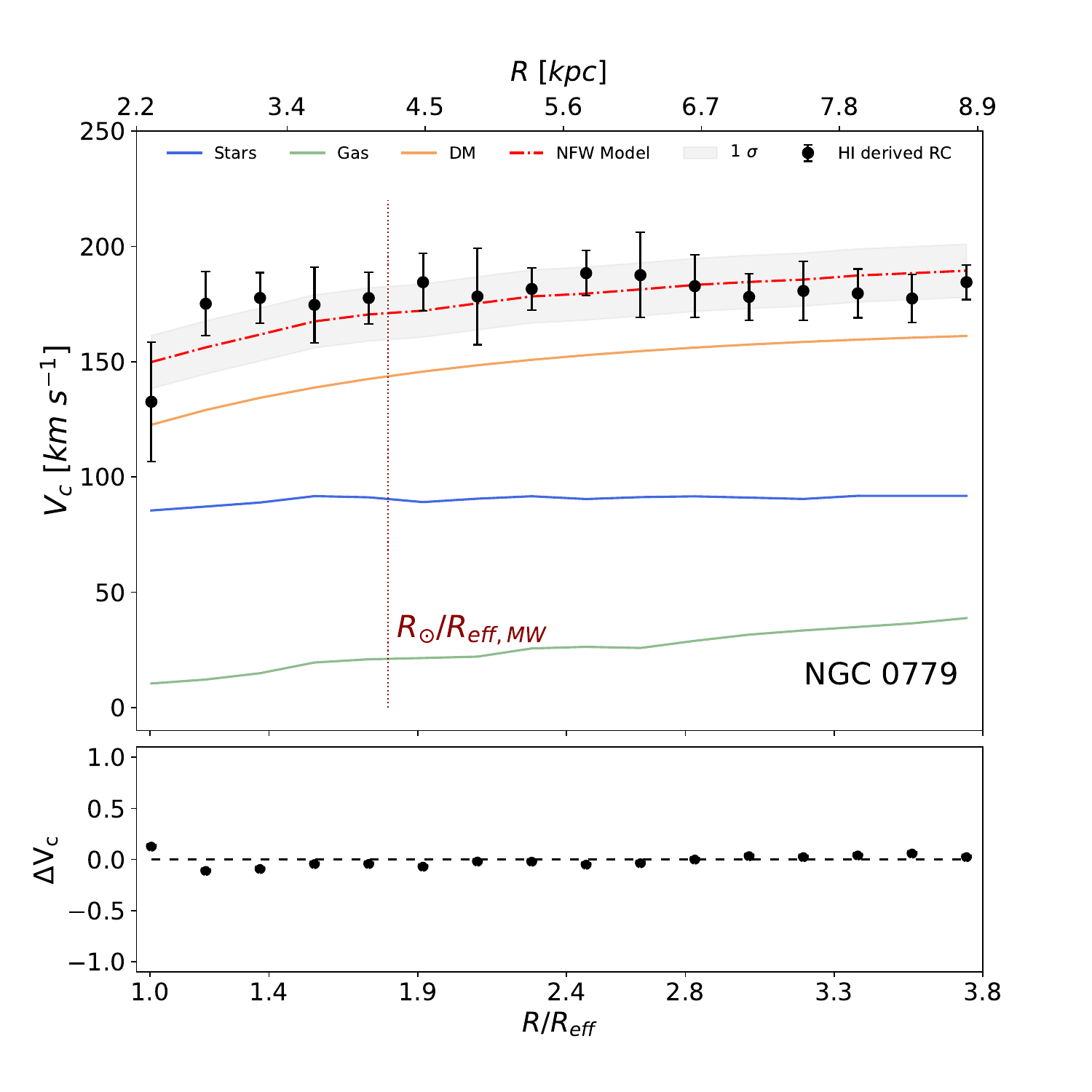}
    \includegraphics[width=0.32\linewidth]{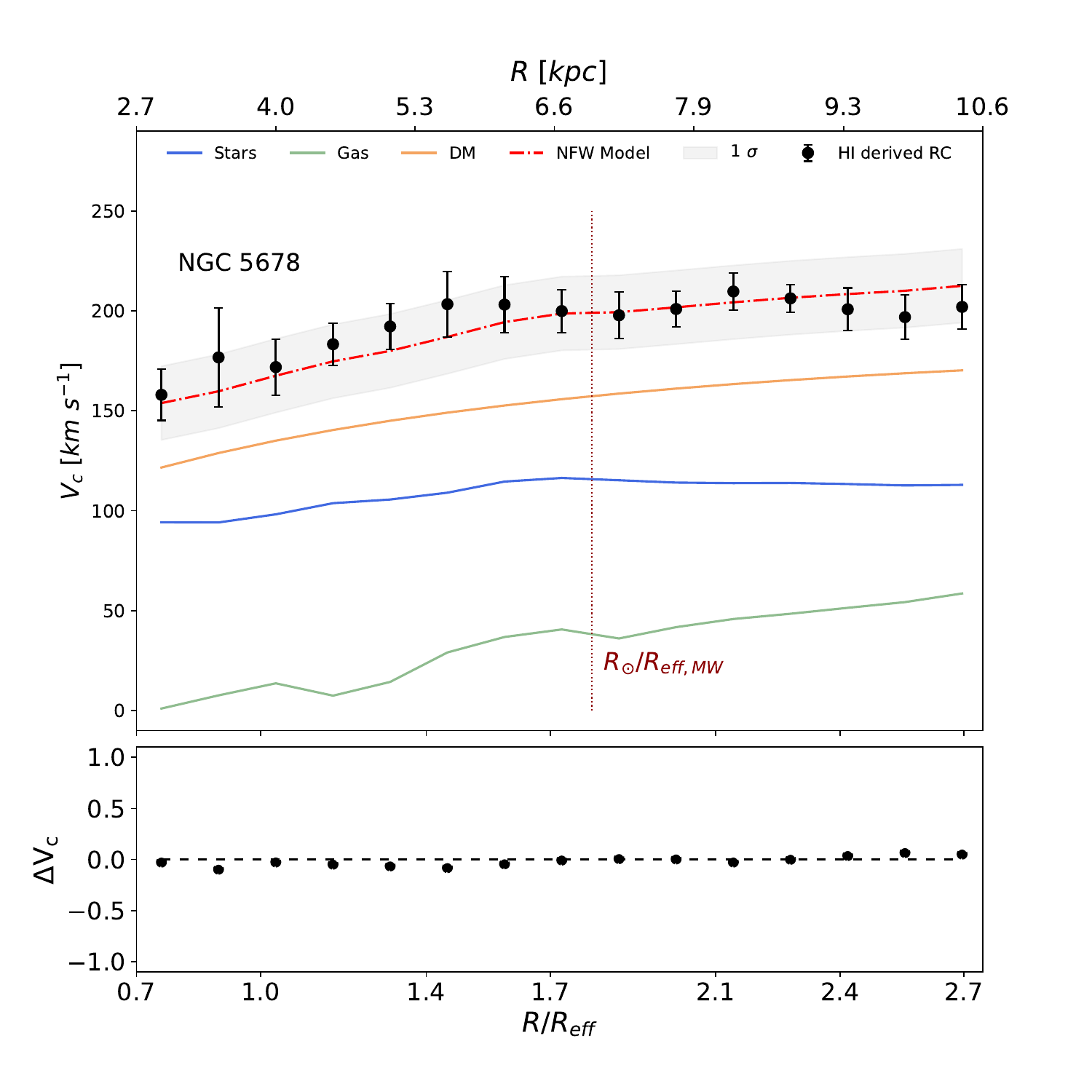}
    \includegraphics[width=0.32\linewidth]{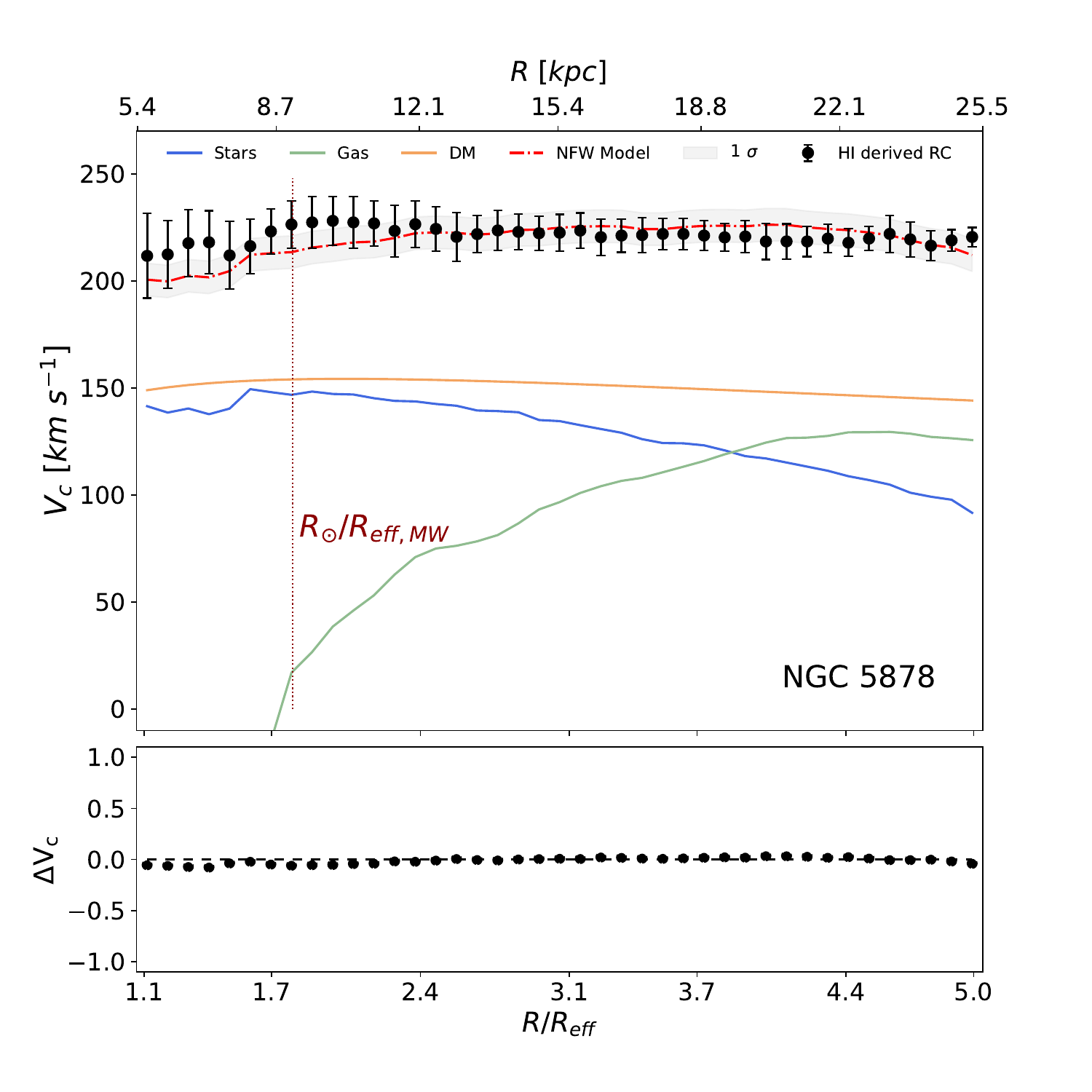}
    \includegraphics[width=0.23\linewidth]{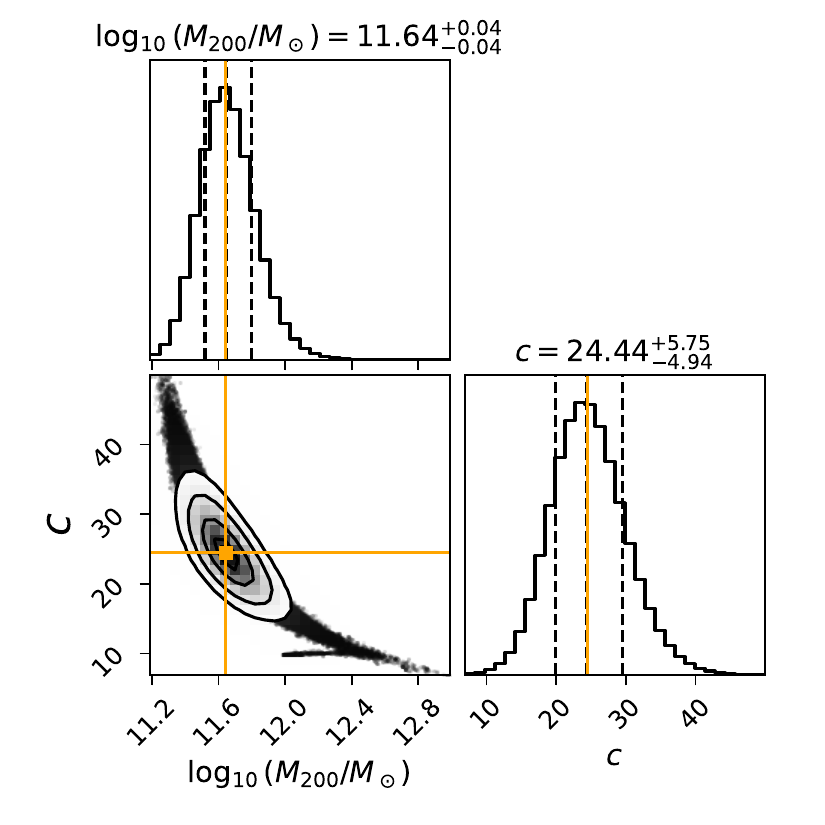}
    \includegraphics[width=0.23\linewidth]{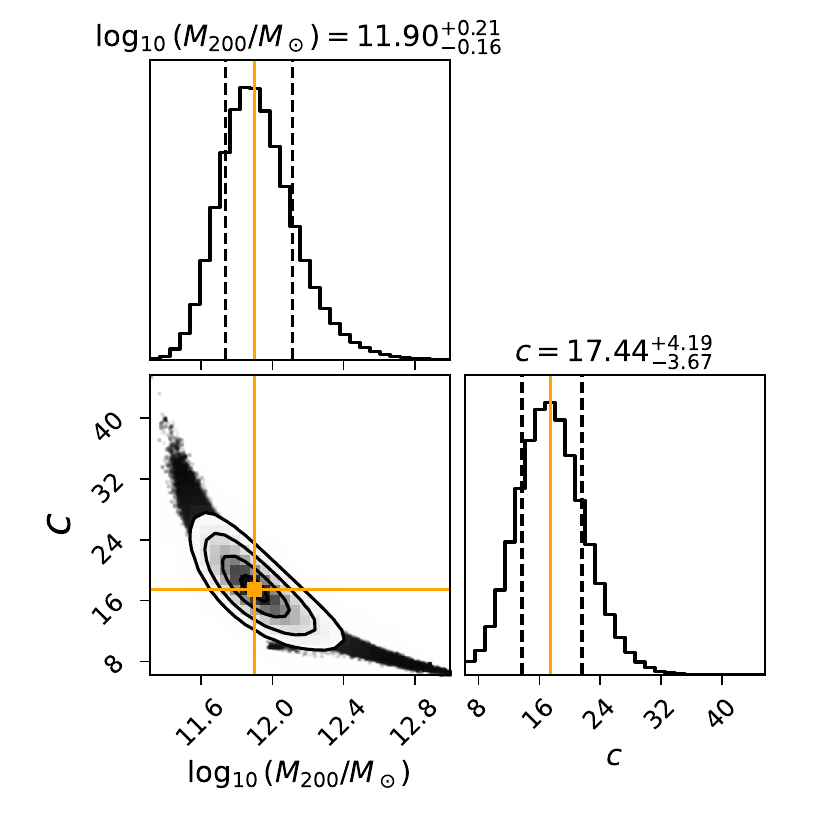}
    \includegraphics[width=0.23\linewidth]{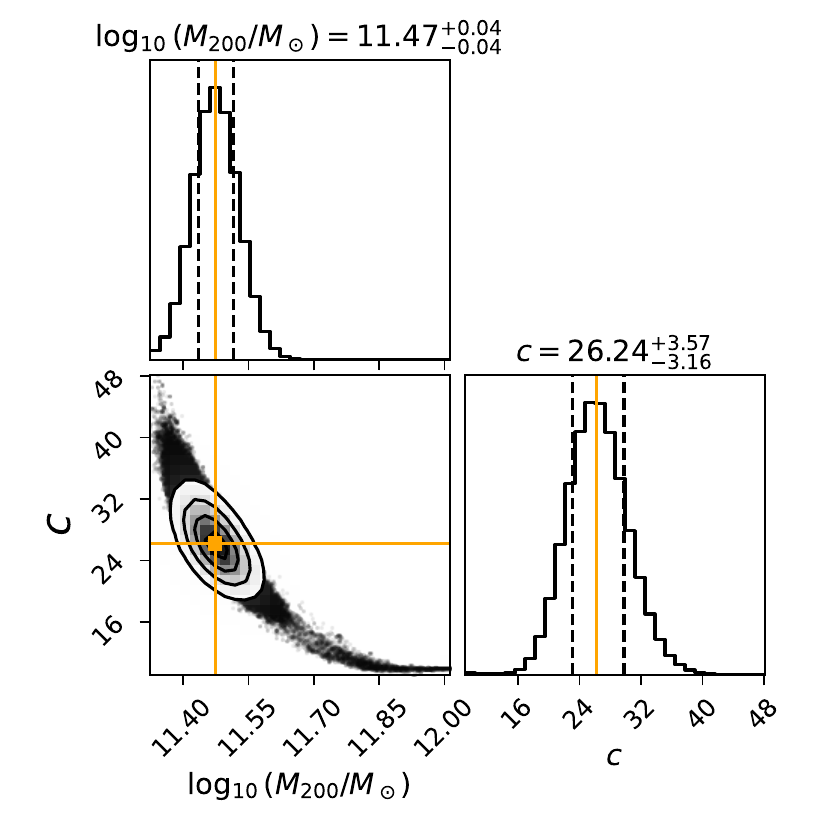}
    \caption{
    Rotation curve (RC) decomposition and best-fit NFW parameters for the observed galaxies. \textit{Rows 1 $\&$ 3:} Decomposed RCs for (a) NGC~7479, (b) NGC~7606, (c) NGC~2775, (d) NGC~0779, (e) NGC~5678, and (f) NGC~5878. Each panel shows the HI-derived circular velocity (black points; $v_{\mathrm{circ}}$), the total best-fit model (red dash-dotted line; $v_{\mathrm{mod}}$), and contributions from gas (green), stars (blue), and the NFW halo (orange). The vertical dark red line marks the equivalent Solar neighborhood at $R = 1.8 \times R_{\mathrm{eff}}$ \citep{licquianewman2015}. The lower panel displays the normalized residuals $(v_{\mathrm{mod}} - v_{\mathrm{obs}})/v_{\mathrm{obs}}$. \textit{Rows 2 $\&$ 4:} Posterior distributions for the best-fit NFW parameters ($M_{200}$, $c$) from our MCMC analysis, corresponding to the RCs directly above them. These results are essential for the stellar-to-halo mass (SHMR) 
    and concentration-mass ($c-M_{200}$) relations discussed in Section~\ref{sec:results}.}
    \label{fig:RCs}
\end{figure*}

The final RC decompositions for the six observed MW analog galaxies in this work are presented in Figure~\ref{fig:RCs}. 
For every system, the modeled circular velocity (based on the NFW DM halo contribution) closely reproduces the observed rotation curves, as evidenced by the small residuals. In most galaxies, the DM component already dominates the gravitational potential at the equivalent solar neighborhood position, whereas the stellar component is dominant in the innermost regions. The only exception is NGC 7479, where the stellar contribution exceeds that of the DM halo out to approximately 3 effective radii\footnote{The effective radius ($R_{\mathrm{eff}}$) is defined as the semi-major axis length of the elliptical isophote that encloses half of the galaxy's total stellar light. We obtained effective radii for all S$^4$G galaxies (Muñoz-Mateus, private communication), based on \citealt{munozmateos2015}
} ($R_{eff}$), after which the DM starts to dominate. The contribution from the gaseous disk is generally minor to the total mass budget, particularly in the central regions where its surface density approaches zero. The velocity profiles derived for gas typically exhibit a smooth rise and maintain a nearly constant value in the outer disks.
NGC~5878 presents a unique case in which the gas component exceeds the stellar contribution at larger radii. This can be explained by its high concentration ($c=26.24$) and lower halo mass ($11.47~log(M_{halo}/M_{\odot})$), as we will discuss in further detail in the next section. Despite the crucial role that extended gas plays in tracing the dynamics and total mass of galaxies, the DM halo and stellar disk remain the primary sources of gravitational potential at the equivalent Solar circle, in agreement with recent work \citep{deisídio2024}.

\subsection{Rotation curves and dark matter profiles in TNG50 MW analogs}

The methodology for deriving kinematics and DM profiles from the TNG50 simulation is more straightforward, as it relies on direct access to the mass distribution of each particle type. This allows us to precisely calculate the gravitational potential and the dynamical contribution of each component (DM, stars, gas) at any point.
Notwithstanding, we must still adopt two key assumptions to infer the circular velocity profile, $v_{\text{circ}}(r)$: 1) the DM halos are assumed to exhibit approximate spherical symmetry and 2) the gravitational force is assumed to be balanced by the centrifugal force: 

\begin{equation}
\frac{m~V^{2}}{r} = \frac{G~M~m}{r^{2}}~\therefore~V_{circ}(r)=\sqrt{\frac{G~M(r)}{r}}
\label{eq:virial}
\end{equation}

Based on these physically motivated conjectures, we perform a mass decomposition and RC analysis on the simulated sample, similar to the methodology applied to our observational data. For each system, we calculate the enclosed mass by summing all particle types within concentric spherical shells of $0.1~\mathrm{kpc}$ width, extending from the galactic center to the outermost bound particle, a radius that can reach several hundred $\mathrm{kpc}$ for the most massive halos.

\begin{figure}
    \centering
    \includegraphics[width=1\linewidth]{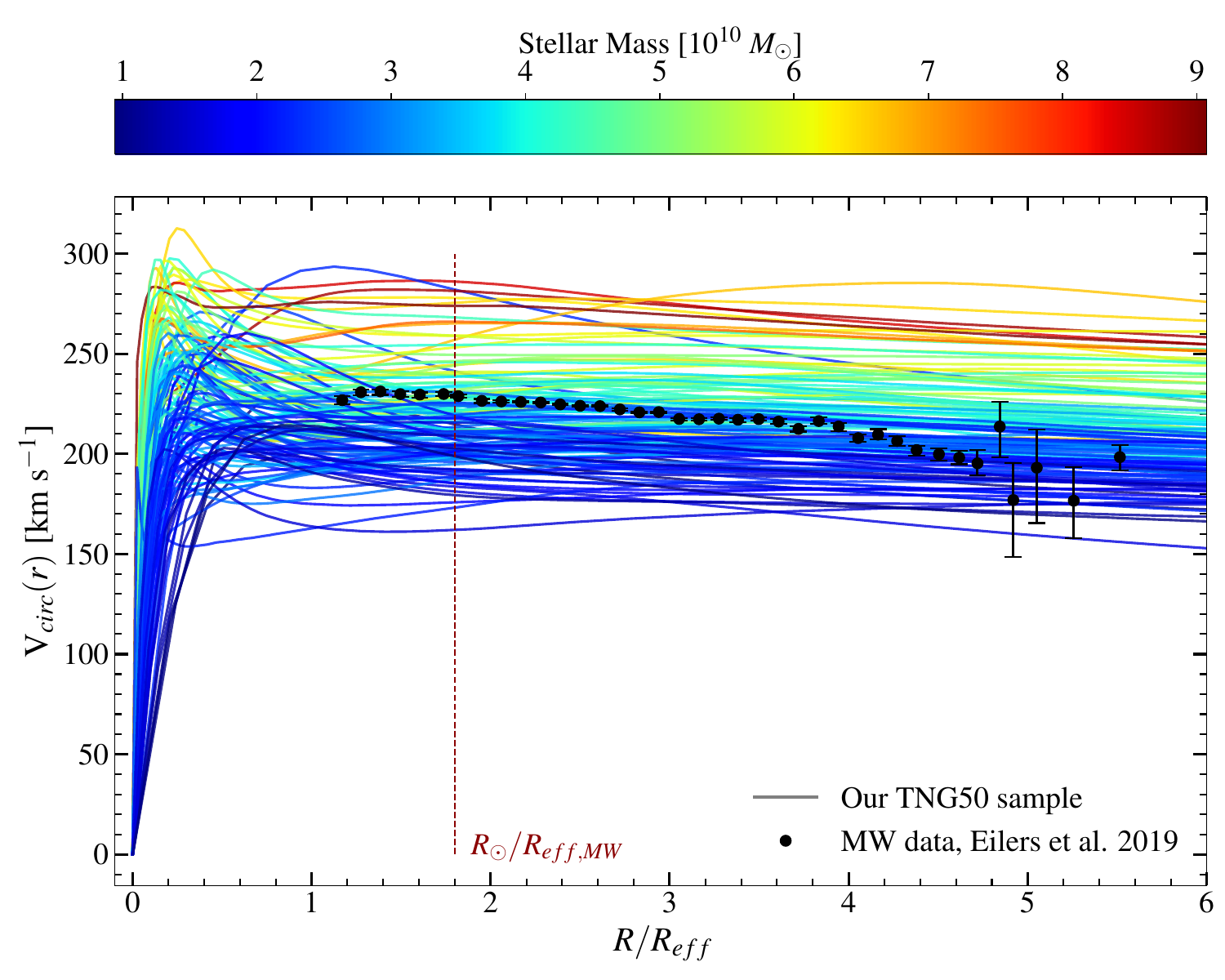}
    \caption{Rotation curves (RCs) for 127 Milky Way (MW) analogs systems from the Illustris TNG-50 cosmological simulation. Curves are color-coded by stellar mass, spanning a range of $1$–$9 \times 10^{10}~M_{\odot}$. We also add the derived RC for the MW in \citet{Eilers_2019}, in black data points. All radii are normalized by the stellar effective radius ($R_{\mathrm{eff}}$), taken directly from the simulation.}
    \label{fig:rc_analogs}
\end{figure}

The circular velocity is then derived from the total enclosed mass, $M(<r)$, using Eq.~\ref{eq:virial}. For our final sample of 127 MW-like galaxies, we generate RCs by quadratically summing the contributions from stars, gas, and DM. 
Figure~\ref{fig:rc_analogs} presents the resulting RCs for the full simulated sample, color-coded by stellar mass. For comparison, we include the high precision RC derived for the MW by \citet{Eilers_2019}. The RCs of the simulated galaxies reproduce the same features as the MW: a steep inner rise followed by a flat outer profile. 
The radius at which the RCs flatten exhibits significant scatter ($\sim0.1-1~R_{eff}$), which is correlated with their halo concentration \citep{Posti2019b}. 
We note that galaxies with lower stellar mass consistently reach their peak circular velocity at lower values. This is an expected trend \citep[e.g.,][]{Posti_2021,Posti2019b}, considering that stellar mass is also an important tracer of the gravitational potential. However, we also identify lower stellar mass galaxies with higher flat rotation velocities. In these cases, a significant contribution from other mass components, such as an extended and massive gas disk can account for the elevated velocities. Consequently, by defining our sample of simulated MW analogs based on maximum rotation velocity, we effectively restrict the halo mass parameter space without imposing a direct limit on stellar mass. The range of colors in Figure~\ref{fig:rc_analogs} reflects this diversity, illustrating the variation in stellar mass that can reside within halos of $M_{200} \gtrsim 10^{12}~M_{\odot}$.

\begin{figure*}
    \centering
    \includegraphics[width=1\linewidth]{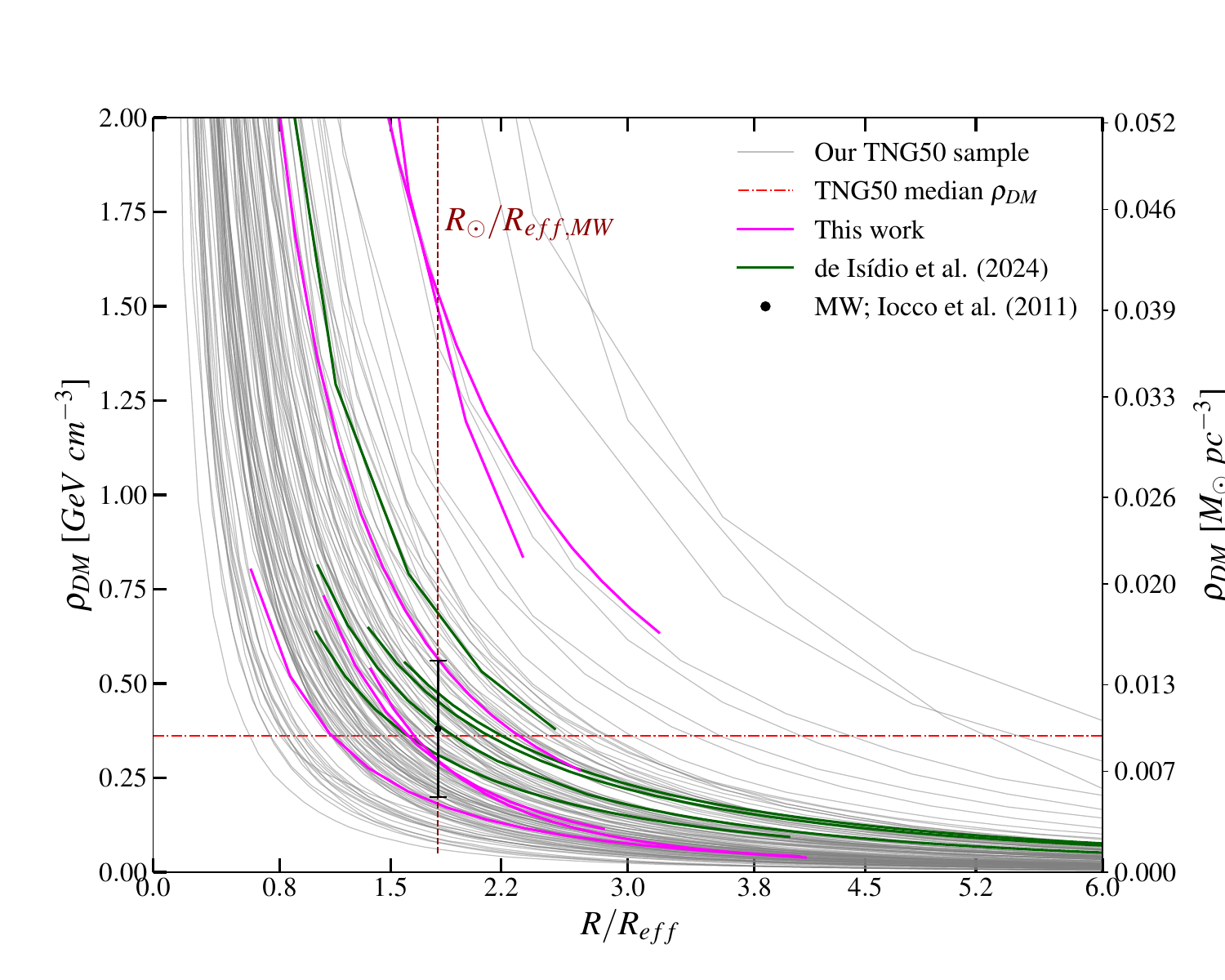}
    \caption{DM density profiles for our sample of 138 MW-like systems. Gray solid lines represent the radial DM density profiles for 127 simulated subhalos from the IllustrisTNG-50 cosmological simulation, calculated in concentric spherical shells of $0.1$ kpc width. Colored solid curves show the DM density profiles derived from the best-fit NFW parameters of our MCMC analysis for the observational sample. The equivalent Solar neighborhood radius, defined as $1.8\times R_{\mathrm{eff}}$, is indicated by the vertical dashed dark red line. The black interval denotes the range of LDMD for the MW from \citet{iocco2011}, based on microlensing and kinematic measurements. The red dash-dotted horizontal line marks the median LDMD ($\rho_{\mathrm{DM}} = 0.39$ GeV cm$^{-3}$) derived from our TNG50 sample.
    For consistency with simulation-based studies \citep[e.g.,][]{pillepich2024}, DM density is also shown in units of $\mathrm{M_{\odot}}~\mathrm{pc}^{-3}$ on a complementary axis.
    }
    \label{fig:density}
\end{figure*}

We derive DM density profiles for the TNG50 galaxies in our sample, based on the the mass enclosed in successive spherical shells, without assuming any specific profile. These profiles are presented in Figure~\ref{fig:density}, alongside the inferred profiles of our observational sample, which comprises the six galaxies from this work (described in Section \ref{subsec:dmhaloprofile}) combined with five systems from \citet{deisídio2024}.

\section{Results and discussion}\label{sec:results}

\subsection{Local dark matter density}\label{subsec:local}

Measuring the DM at the Solar neighborhood is a very challenging task. This difficulty arises not only from our embedded position within the system, which introduces strong observational biases \citep{licquianewman2015}, but also from the numerous assumptions and approximations required to derive this quantity \citep{Catena_2010,Salucci_2010,reid2014}. We use external galaxies with similar DM halos in an effort to draw analogies with our own. By deriving the DM density radial profiles in these MW analogs, we can estimate the typical dark-matter density range at the Solar-equivalent radius.

The DM density radial profiles that we derived for our sample of observed and simulated MW analogs are presented in Figure~\ref{fig:density}. To enable a direct comparison between our sample of analog galaxies and the MW, we parametrize galactocentric distances in terms of the galaxy's stellar effective radius ($R_{\mathrm{eff}}$). Adopting a disk scale length for the MW of $2.6~\mathrm{kpc}$, we obtain an effective radius of $R_{\mathrm{eff, MW}} = 4.5~\mathrm{kpc}$ \citep{licquianewman2015,Zhou_2023}. The Solar radius can thus be expressed as $R_{\odot} \approx 1.8~\times R_{\mathrm{eff, MW}}$. With S4G effective radii available for our observed sample, we show in Figure~\ref{fig:density} the DM radial density profiles of our sample in units of Reff, allowing for an immediate appreciation of the range in DM density values that our sample galaxies exhibit at the galactocentric distance that corresponds to the Sun's location in the MW. For the TNG50 sample of MW analogs, we do not have available effective radii attached to light distributions. However, considering that age, metallicity and extinction radial gradients have very little impact in the mid-IR emission \citep{munozmateos2015}, we use the half-mass radius provided by the TNG50 database as an equivalent radial parametrization to the one we use for our observed sample. We show the median LDMD derived from our TNG50 sample in Figure \ref{fig:density}. This value shows excellent agreement with the range estimated for the MW in \citet{iocco2011}.

We note that the MW's bar, with semi-major axis of ~4 kpc (corresponding to ~0.9 Reff), points to a lack of spherical symmetry in these inner regions. However, the region of interest in this study is the radial position of the Sun in the MW and the resulting DM profiles in our study cover the radial range between 1Reff and 2-5 Reff, thereby excluding potentially complex bar-dominated regions.

Figure \ref{fig:density} shows that the TNG50 simulated sample displays a broader range in DM density values at the equivalent Solar location, encompassing both higher and lower values than those seen in observations. We attribute this difference primarily to the size differences between the two samples: the observed sample, by the challenging nature of the observations, is quite limited, whereas the TNG50 simulation provides a diverse population of over a hundred systems with matched mass, kinematics, and morphological properties. The DM density profiles derived in \citet{deisídio2024} show good agreement with that of four galaxies observed in this work and are consistent with the overall distribution of profiles from our simulated galaxies.

Within the TNG50 sample, we identify 8 systems with particularly high DM content that nonetheless satisfy our selection criteria for MW-like galaxies. Although these systems could potentially be distinguished from the rest based on their compactness or their low stellar mass (i.e., ($\log(M_{\star}/M_{\odot}) < 10$), we note that their DM profiles share similarities with the 2 outliers from our observed sample. In total, these 10 systems (8 simulated, 2 observed) represent the most significant outliers in our analysis, characterized by their compact size, lower rotational support and, for the observed ones, their proximity ($\lesssim 17~Mpc$). 

\begin{figure}
    \centering
    \includegraphics[width=1\linewidth]{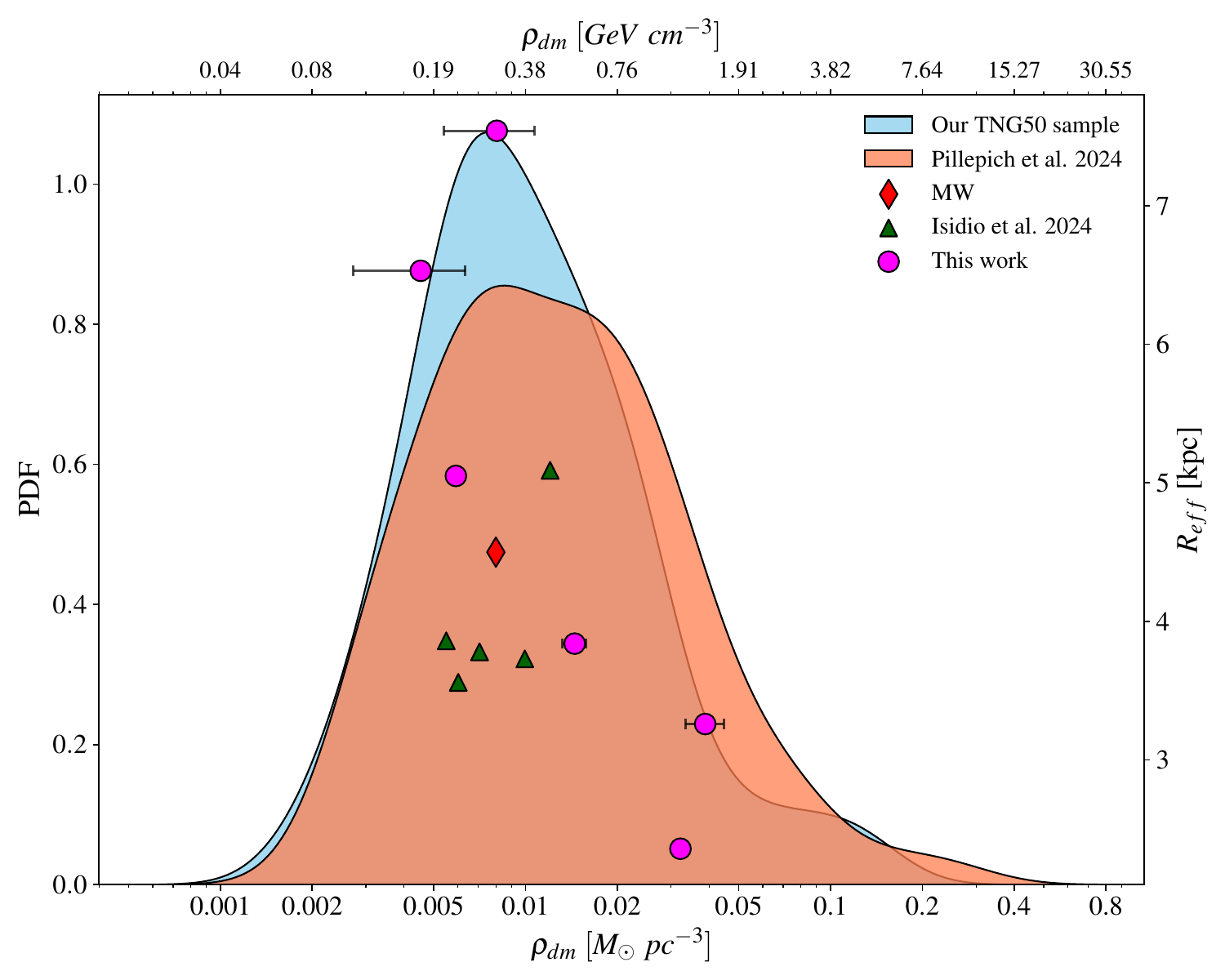}
    \caption{
    Comparison of the LDMD derived for our sample of 127 simulated MW analog galaxies (in sky blue) with values from the 130 TNG50 analogs studied in \citet{pillepich2024} (coral), under a different selection criteria. In both samples we apply the equivalent Solar neighborhood definition detailed in Section~\ref{subsec:local}. We include LDMD estimates for our observed galaxies (pink data points), the sample from \citet{deisídio2024} (dark green triangles), and the estimate for the MW predicted by the Standard Halo Model ($\rho_{DM}=0.3~GeV~cm^{-3}$; red diamond), positioned according to their effective radius $R_{\mathrm{eff}}$.}
    \label{fig:local}
\end{figure}

Figure~\ref{fig:local} shows the distribution of the LDMD calculated at the equivalent Solar position within our simulated sample (127 subhalos), compared to that derived by \citet{pillepich2024}  for their MW analog sample (130 subhalos) using our definition of the solar position. Our selection criteria yield a narrower LDMD distribution, peaking within the range reported by most previous MW studies: $0.34~\mathrm{GeV}~\mathrm{cm}^{-3}$ (equivalent to $\approx ~0.009~M_{\odot}~\mathrm{pc}^{-3}$; see Table~\ref{tab:LDMD} for a compilation of references). We also include data points derived from our observational analysis, vertically positioned according to their $R_{\mathrm{eff}}$, with error bars estimated from our MCMC analysis. We find that systems with the smallest $R_{\mathrm{eff}}$ (NGC 2775, $R_{eff} = 3.26~kpc$; NGC 0779, $R_{eff}=2.35~kpc$) exhibit the most extreme LDMD values ($\rho_{DM,\odot}=1.770$ and $1.230$, respectively). This trend is directly linked to the halo concentration parameter, where systems with higher concentrations are more DM-dominated, which consequently results in a higher inferred density at the equivalent Solar neighborhood radius. 

While these compact systems are included in the full analysis, we exclude them for the purpose of estimating the final LDMD range, avoiding potential bias from these outliers. Our resulting LDMD estimate $0.17 - 0.46~GeV~cm^{-3}$ is therefore based on four galaxies from this work (NGC~7479, NGC~7606, NGC~5678, NGC~5878), five from \citet{deisídio2024} (NGC~2903, NGC~3521, NGC~4579, NGC~4698, NGC~5055), and 127 TNG50 systems. This constrained range is consistent with most previous estimates for the MW and its analogs (see references in Table~\ref{tab:LDMD})

\begin{deluxetable}{lcc}
\centering
\tabletypesize{\footnotesize}
\tablewidth{0pt}
\tablecaption{Comparison of LDMD estimates in the literature. The last column indicates the technique used to derive the LDMD estimates. Method abbreviations: Theo = theoretical; CE = centrifugal equilibrium; GM = global modeling; ML = microlensing; VM = vertical motion; SIM = simulations; PT = Pulsar timing measurements; SS = stellar streams. The asterisk symbol denotes estimates derived from MW analog sample analysis. A useful conversion is $1~M_{\odot}~\mathrm{pc}^{-3} = 37.96~\mathrm{GeV}~\mathrm{cm}^{-3}$.
\label{tab:LDMD}}
\tablehead{
\colhead{} & \colhead{LDMD} & \colhead{} \\
\cline{2-2}
\colhead{Authors} & \colhead{$\textrm{GeV}~\textrm{cm}^{-3}$} & \colhead{Method} \\
}
\startdata
\textit{Standard Halo Model} & 0.30 & Theo \\
\cite{Catena_2010} & $0.39 \pm 0.02$ & GM \\
\cite{Salucci_2010} & 0.43 $\pm 0.11$ & CE \\
\cite{Weber_2010} & 0.20 - 0.40 & GM \\
\cite{iocco2011} & 0.20 - 0.56 & ML \& GM \\
\cite{Bovy2012} & $0.30 \pm 0.10$ & VM \\
\cite{zhang2013} & $0.25 \pm 0.09$ & VM \\
\cite{Nesti_2013} & $0.47 \pm 0.05$ & GM \\
\cite{Bienayme2014} & $0.54\pm0.04$ & VM \\
\cite{Sofue2015} & $0.26 \pm 0.11 $ & GM \\
\cite{PatoDynamical_2015} & $0.42 \pm 0.02$ & GM \\
\cite{huang2016} & $0.32 \pm 0.02 $ & GM \\
\cite{McMillan2017} & $0.40 \pm 0.04$ & GM \& SIM \\
\cite{Eilers_2019} & $0.30 \pm 0.03$ & GM \\
\cite{Karukes_2020} & 0.45 - 0.49 & GM \\
\cite{Chakrabarti_2021} & $0.13\pm1.32$ & PT \\
\cite{Palau_2023} & $0.23 \pm0.02$ & SS \& GM \\
\cite{pillepich2024} & 0.20 - 0.50 & SIM* \\
\cite{deisídio2024} & 0.21 - 0.55 & GM* \\
\textbf{This work} & \textbf{0.17 - 0.46} & \textbf{GM* \& SIM*} \\
\enddata
\end{deluxetable}

\subsection{Galaxy-Halo connection}\label{subsec:galaxy-halo}

The intrinsic shape of a DM halo is a topic explored by many authors \citep{Warren1992,Governato_2012,Palau_2023}. Seminal works based on DM-only simulations typically found prolate halos \citep{Frenk1988}, while later simulations that incorporated baryonic physics such as gas cooling \citep{Dubinski1994}, star formation \citep{Bryan2013,Di_Cintio_2013}, AGN and stellar feedback \citep{auriga2017,Nelson_2019} often produced more oblate, flattened distributions. The precise dynamical mechanisms responsible for this transformation, and the extent to which baryons can alter the inner density profile of a halo, are not yet fully understood. Nevertheless, it is evident that there exists a strong connection between the luminous component and the DM halo structure \citep{wechslertinker2018}.

Given the complex nature of galaxy formation, an effective way to address the galaxy-halo connection is to analyze scaling relations that link both components, that is, the stellar and halo masses. The abundance matching technique presented in \cite{Moster_2010} -- see \citealt{wechslertinker2018} for a recent review -- yields a stellar-to-halo-mass relatin (SHMR) characterized by a steep power-law slope at low masses, which gradually flattens near a characteristic halo mass of $M_{200} \sim 10^{12}~M_{\odot}$ and stellar mass $M_{\star} \sim 5 \times 10^{10}~M_{\odot}$ ($f_{\star} \sim20\%$). The physical interpretation of this transition and the precise location of different galaxy types (e.g., dwarfs, spirals, ellipticals) within this framework remain subjects of active debate in the literature. Recent works using spatial and spectral high-resolution data also derive such trend \citep{Posti2019a,Posti2019b,Posti_2021,mancera2022, DiTeodoro2022}, following previous predictions in the literature \citep[e.g.,][]{Moster_2010,Moster_2012, Roduiguez-puebla2015}.  In this context, the star formation efficiency ($f_{\star}$) and the baryon fraction ($f_{\mathrm{b}}$) within our sample provide crucial insights into the remaining gas reservoir (the fuel available for future star formation) and offer valuable constraints on the fraction of ``missing baryons" on galactic scales.

\begin{figure}
    \centering
    \includegraphics[width=1\linewidth]{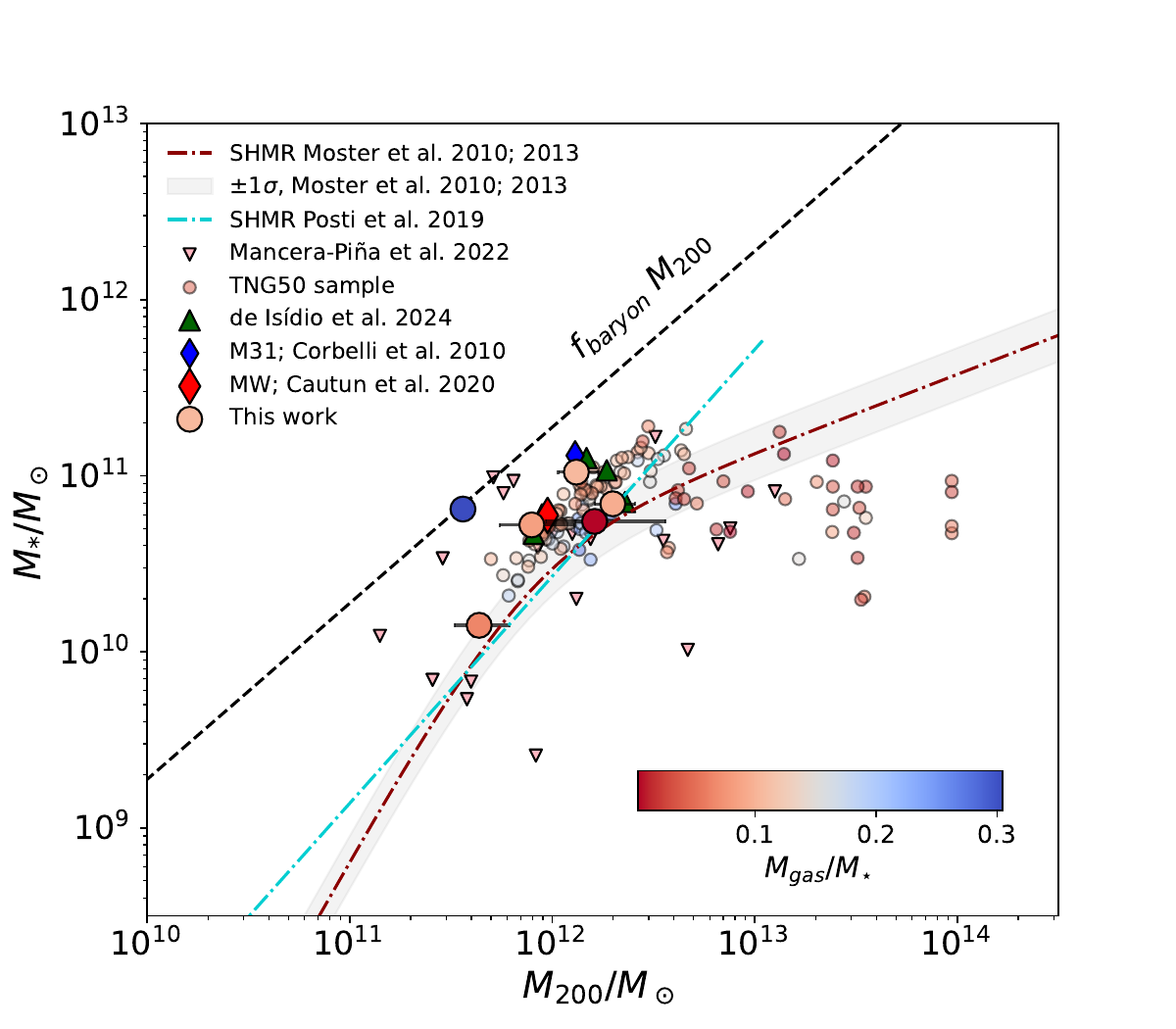}
    \includegraphics[width=1\linewidth]{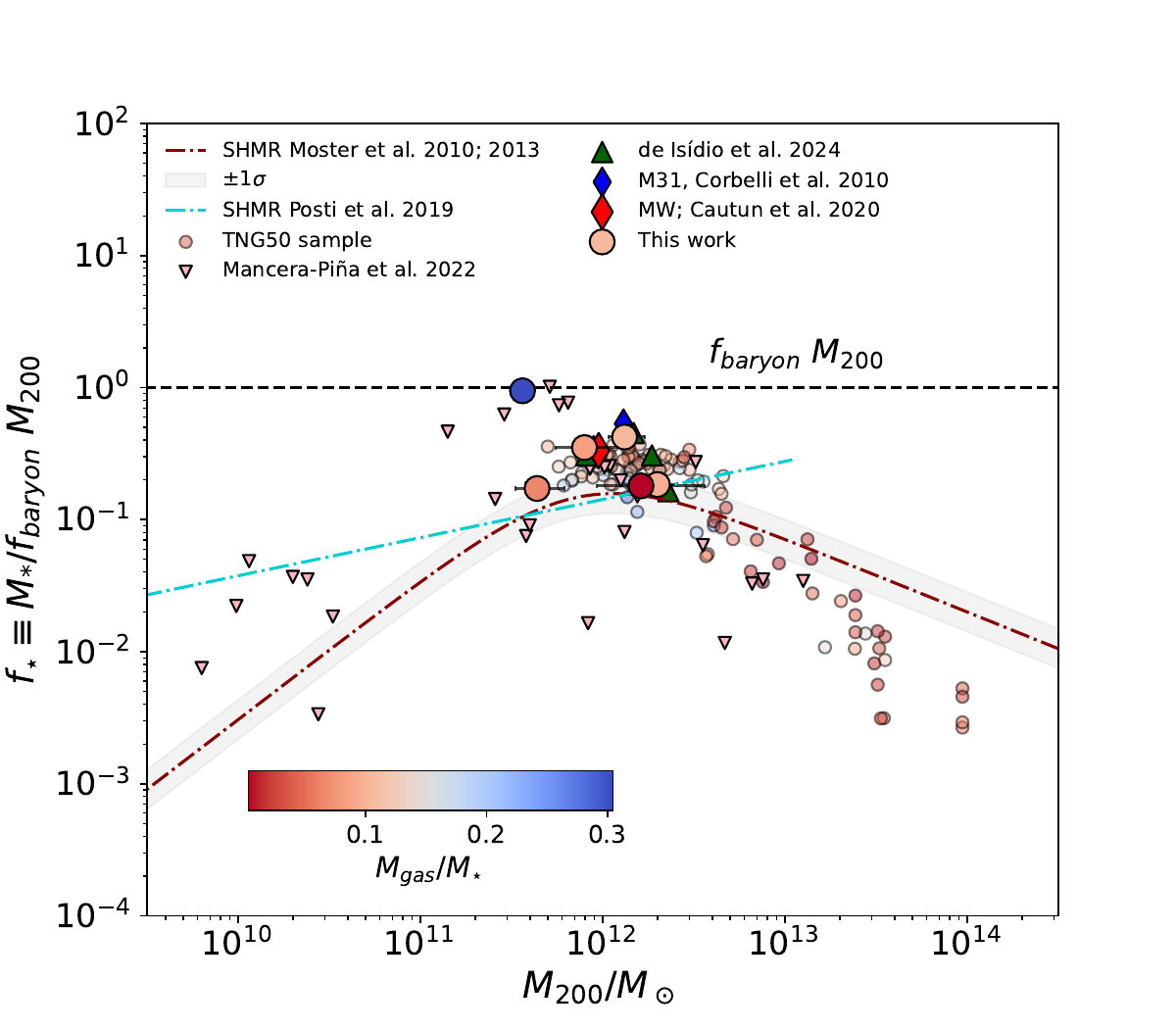}
    \caption{Top panel: Stellar-to-halo mass relation, SHMR. Bottom panel: Star formation efficiency, $f_{\star} \equiv M_{\star}/f_{baryon}M_{200}$.
    The dash-dotted dark red curve shows the SHMR from \citet{Moster_2010}, with its $1\sigma$ scatter. The teal line represents the linear SHMR we derived from a SPARC sample \citep{Lelli_2016} analyzed in \citep{Posti_2019}. Light pink points correspond to data from \citet{mancera2022}, while green triangles denote galaxies from the VIVA \citep{chung_2009} and THINGS \citep{Walter_2008} surveys studied in \citet{deisídio2024}. The dashed black line indicates the universal baryon fraction ($f_{\mathrm{b}} = \Omega_{\mathrm{b}}/\Omega_{\mathrm{m}}$), representing the theoretical upper limit for the baryonic mass within a halo.}
    \label{fig:SHMR}
\end{figure}

In Figure~\ref{fig:SHMR}, we present the derived stellar mass ($M_*$) as a function of halo mass ($M_{200}$) for both the observational and the simulated TNG50 sample. For context, we include the SHMR from \citet{Moster_2010} and the linear fit (in log–log space) using \citet{Posti_2019} mass estimates from 110 disc galaxies in the SPARC database \citep{Lelli_2016}. Our entire observed sample and the majority ($>50\%$) of TNG50 galaxies are concentrated above the $\approx 12~log(M_{200}/M_{\odot})$ peak from \citet{Moster_2010} relation, and below the cosmological baryonic mass limit ($f_{baryon} M_{200} = 0.188 $, \citealt{planck}). 
The data points from \citet{mancera2022}, exhibit significantly greater scatter than ours. This is a direct result of their sample composition, which includes both massive spirals and dwarf galaxies (22 and 10, respectively). Dwarf galaxies are typically DM-dominated \citep[e.g.,][]{Oman_2015} and consequently occupy the lower stellar mass region ($\log(M_*/M_\odot) \lesssim 9$)  of the parameter space; this accounts for the broader distribution of datapoints from \citet{mancera2022}. As we focus this work on MW analogs, which are more massive late-type galaxies, this dwarf galaxy locus is not covered by our sample. Galaxies from VIVA \citep{chung_2009} and THINGS \citep{Walter_2008} surveys studied in \cite{deisídio2024}, are in strong agreement with our results.

We also show the stellar mass fraction normalized by the average cosmological baryon fraction in Figure \ref{fig:SHMR}. The y-axis, $f_{\star} \equiv M_{\star}/(f_{baryon}M_{200})$, represents the efficiency with which baryons have been converted into stars. 
Galaxies with higher $f_{\star}$ values are those that are more stellar dominated.
Half of our observational sample (NGC 7606, NGC 2775, NGC 0779) lies within the $1\sigma$ scatter of the \citet{Moster_2010} relation, consistent with the expected peak in galaxy formation efficiency \citep{Silk}. The other half (NGC 7479, NGC 5678, NGC 5878) lies systematically above this relation. This suggests that these systems have, in fact, higher stellar mass fractions than predicted by \citealt{moster2010}.
The more massive halos in this subgroup appear to align with the relation by \citet{Posti_2021}, where late-type galaxies may follow a monotonically rising SHMR, distinct from the declining trend seen in early-type systems. The isolated position of NGC~5878, a less massive spiral, in Figure~\ref{fig:SHMR} may indicate a distinct evolutionary pathway characterized by exceptionally high star formation efficiency.

While our sample has a lower mass limit that could potentially bias a full analysis of the SHMR shape, the region of primary interest, $M_{\mathrm{halo}} \sim 10^{12}~M_{\odot}$, is well-represented.
The physical mechanisms responsible for the declining star formation efficiency at both the low- and high-mass ends of the SHMR are primarily linked to feedback processes. At the low-mass end, energetic feedback from young stars can effectively disrupt the interstellar medium and outweigh star formation \citep[][and references therein]{Veilleux}. At the high-mass end, feedback from active galactic nuclei (AGN) plays a critical role in driving powerful outflows \citep[e.g.,][]{Weinberger_2016,Harrison}. In both regimes, these feedback processes suppress the inflow and delay or prevent gas cooling, thereby promoting galactic quenching \citep{Posti_2021}.
A reliable estimate of the gas mass (even as a lower limit, given the inherent loss of large-scale, diffuse emission in interferometric data) provides crucial insight into a galaxy's fuel reservoir and its star formation history. We analyze the gas-to-stellar mass ratio ($M_{\mathrm{gas}}/M_{\star}$), represented by the color scale in Figure~\ref{fig:SHMR}.
Despite the modest size of our observed sample, we identify a clear bimodality: gas-rich (blue) galaxies with $M_{\mathrm{gas}}/M_{\star} \gtrsim 10\%$, and gas-poor (red) galaxies with $M_{\mathrm{gas}}/M_{\star} \lesssim 5\%$ residing in DM halos of different masses. Figure \ref{fig:SHMR} clearly shows a DM halo mass trend spanning from the most gas-rich to the less gas-rich system in our sample. 
This trend is consistent with the findings of \citet{Cui2021}, who showed that at a fixed stellar mass, bluer galaxies tend to reside in less massive halos, while redder galaxies inhabit more massive ones. 
Conversely, at a fixed halo mass, bluer galaxies tend to have formed stars more efficiently, resulting in higher stellar masses, whereas redder galaxies typically exhibit lower stellar masses for the same halo mass. 
A complete explanation for the scatter in the SHMR requires detailed knowledge of individual halo assembly histories \citep[e.g.,][]{Artale2019,Cui2021}, which is beyond the scope of this work.

\section{Conclusions} \label{sec:conclusion}

In this work we study a combined sample of 11 observed and 127 simulated MW analog galaxies to understand how DM is distributed in these systems. For the observational sample, we apply a full dynamical treatment to decompose the mass components within their halos; we use the \textsc{$^{3D}$Barolo} algorithm to derive RCs by combining HI kinematics with mid-infrared imaging. This approach allows us to robustly account for the gravitational contributions of stars and gas, thereby isolating the DM component. Through a careful decomposition of these main galactic components, we show our derived RCs and the best-fit NFW values for concentration and M200 to model these galaxies' DM halos. The TNG50 simulated sample provides a diverse and statistically significant suite of analogs that allows us to compare state-of-the-art theoretical and empirical prescriptions with real data.

Our main results are summarized below.
\begin{enumerate}
    \item[(i)]  The NFW profile reproduces the HI-derived RCs of all observed MW–analog galaxies in our sample. Although the TNG50 DM density profiles show larger object–to–object scatter — reflecting the broader diversity of halo properties in a cosmological volume — their shapes remain consistent with NFW. This agreement across both data and simulations supports a working universality in the halo structure of MW analogs at galactic radii.
    \item[(ii)]  The inferred LDMD for our MW–analog sample falls in the range $0.17-0.46~GeV~cm^{-3}$ ($\approx 0.004-0.012~M_{\odot}~pc^{-3}$), consistent with MW values from the literature. The narrowness and MW consistency of this window—even for the most DM–dominated, compact halos—further points to an underlying universality in the inner halo of MW–like systems.
    \item[(iii)] Based on our derived DM, stellar and gas mass estimates ($M_{200}$, $M_{\star}$, $M_{\mathrm{gas}}$) for the combined sample of 138 MW analogs, we examine the stellar–to–halo mass relation (SHMR) and find a clear bimodality: at fixed $M_\star$, gas–rich (blue) galaxies reside in less massive halos than gas–poor (red) systems; at fixed $M_{200}$, blue galaxies have higher $M_\star$ than red ones. This indicates substantial diversity in baryonic growth histories even among halos of comparable mass.
\end{enumerate}

This work demonstrates the potential of analyzing HI kinematics in nearby galaxies to overcome the challenges of studying the DM halo of the MW from within. Next generation radio telescopes, such as the ngVLA and SKA, will enable detailed kinematic studies of cold gas disks in thousands of galaxies, both nearby and at high redshift. Their unprecedented sensitivity and resolution will provide a deeper understanding of the connection between galaxies and their halos. This refined view of DM distribution on galactic scales will, in turn, deliver crucial astrophysical context for interpreting the results of upcoming direct detection experiments on Earth.

\begin{acknowledgments}

M.C.C.S. acknowledges the support of the \textit{Brazilian National Research Council} (CNPq, Brazil). K.M.D. acknowledges the support of the Serrapilheira Institute (grant Serra-1709-17357) as well as that of the Brazilian National Research Council (CNPq grant 308584/2022-8) and of the Rio de Janeiro State Research Foundation (FAPERJ grant E-26/ 200.952/2022), Brazil. T.S.G. would also like to thank the support of CNPq (Productivity in Research grant 314747/ 2020-6) and the FAPERJ (Young Scientist of Our State grant E-26/201.309/2021). 
DCR thanks \textit{Centro Brasileiro de Pesquisas Físicas} (CBPF) for hospitality, where part of this work was done. He also acknowledges CNPq, \textit{Fundação de Amparo à Pesquisa e Inovação do Espírito Santo} (FAPES, Brazil) and \textit{Fundação de Apoio ao Desenvolvimento da Computação Científica} (FACC, Brazil) for partial support.

\end{acknowledgments}

\bibliographystyle{mnras}
\bibliography{Bibliography} 

\end{document}